\documentclass[twocol]{ametsoc}

\usepackage{xr}
\usepackage{booktabs} \newcommand{\ra}[1]{\renewcommand{\arraystretch}{#1}}
\usepackage{lineno}

\journal{jcli}

\bibpunct{(}{)}{;}{a}{}{,}

\title{Response of Vertical Velocities in Extratropical Precipitation Extremes to Climate Change}

\authors{Ziwei Li\correspondingauthor{Department of Earth, Atmospheric,
and Planetary Sciences, Massachusetts Institute of Technology, 77 
Massachusetts Ave., Cambridge, MA 02139.} 
and Paul O'Gorman
}

\affiliation{Department of Earth, Atmospheric and 
Planetary Sciences, Massachusetts Institute of Technology, Cambridge, MA, USA}

\email{ziweili@mit.edu}

\begin{document}

\abstract{ Precipitation extremes intensify in most
regions in climate-model projections.  Changes in vertical velocities 
contribute to the changes in intensity of precipitation extremes but remain poorly understood.  Here, we find that mid-tropospheric vertical velocities in extratropical precipitation extremes strengthen overall in
simulations of 21st-century climate change. For each extreme event, we solve
the quasi-geostrophic omega equation to decompose this strengthening 
into different physical contributions. We first consider a dry decomposition
in which latent heating is treated as an external forcing of upward motion. 
Much of the positive contribution to upward motion from increased latent heating is offset by negative contributions
from increases in dry static stability and changes in the horizontal length
scale of vertical velocities.  However, taking changes in latent heating as given is a limitation when the aim is to understand changes in precipitation, since
latent heating and precipitation are closely linked. 
Therefore, we also perform a moist decomposition of the
changes in vertical velocities in which latent heating is represented through a
moist static stability. 
In the moist decomposition, changes in moist static stability play a key role and contributions from other factors such as changes in the depth of the upward motion increase in importance.
While both dry and moist decompositions are self-consistent, the
moist dynamical perspective has greater potential to give insights into 
the causes of the dynamical contributions to changes in precipitation extremes in different regions.  }

\maketitle

%

\section{Introduction}

Projected changes in the intensity of precipitation extremes in response to
climate warming may be decomposed into a positive thermodynamic contribution
(roughly 6\% K$^{-1}$ in the extratropics) from increased humidity and a
dynamical contribution from changes in vertical velocities \citep{Emori2005,
OGorman2009}.  The dynamical contribution is responsible for most of the
geographical and seasonal variation of the projected response of precipitation
extremes, and it is large enough to cause decreases in the intensity of
precipitation extremes over parts of the subtropical oceans \citep{Pfahl2017}.
Here we focus on the dynamical contribution in the extratropics which is
relatively robust across coupled general circulation models (GCMs)
\citep{Pfahl2017} but remains challenging to understand given the importance of
latent heat release
in extreme precipitation events \citep{Nie2018}. 

\citet{Dwyer2017} identified extreme precipitation events in coupled GCM
simulations using a high percentile of the 3-hourly precipitation rate and then
calculated the spatial extent and duration of the events based on a minimum
threshold of 25\% of that percentile.  
Extreme precipitation events in these simulations were estimated to be
of order 700km in horizontal
extent and 14 hours in duration.  As a result, the quasi-geostrophic omega
(QG-$\omega$) equation is a useful approximate tool to better understand these
events \citep[e.g., ][]{OGorman2015,Nie2018}.  According to the QG-$\omega$
equation, ascent is forced by large-scale balanced flow and a
feedback from diabatic heating \citep{Nie2015}.  \citet{Tandon2018} (hereafter
T18) and \citet{Tandon2018a} performed scaling analyses of the terms in the
QG-$\omega$ equation for extreme precipitation events in GCM simulations under
climate change.  Here, we take an important further step by numerically solving
the QG-$\omega$ equation in domains centered on such events.

We decompose the projected changes in vertical velocities into different
physical contributions. We begin with a dry decomposition in which diabatic
heating (dominated by latent heating) is treated as an external forcing.
Consistent with the analysis of T18, we find that increased diabatic heating
tends to amplify the changes in vertical velocities, and that this is partially
offset by the weakening effects of increased dry static stability.
However, we find that the contribution of changes in horizontal length scale is 
relatively unimportant in the subtropics in contrast to the 
analysis of T18 as discussed in detail in Appendix C. 

The dry decomposition is useful as a first step and follows the approach used
in previous work, but it treats diabatic forcing
as an external forcing when it is really
part of the internal dynamics of the atmosphere \citep{Emanuel1994}.  In
particular,
changes in latent heating should not be taken as independent from the vertical motion
if the aim is to understand changes in precipitation since the surface precipitation rate is
closely related to the column-integrated latent heating.  To mitigate this
problem, we also introduce a moist dynamical decomposition of changes in
vertical velocities in which latent heating is represented using a moist static stability.  The reduction from the dry static stability to the moist static
stability may be thought of as equivalent to the latent heating feedback discussed previously \citep{Nie2015,Nie2018}.
In the moist decomposition, factors such as changes in the moist static stability and 
increases in the depth of the upward motion 
in each event play an important role.

We first describe the simulations and output used and our definition of extreme precipitation events (section \ref{sec:data}). 
We then describe the numerical inversion of the QG-$\omega$ equation in the extreme precipitation events (section \ref{sec:qgomega}), 
the response of the vertical velocities to climate change (section \ref{sec:omega_vs_omega_QG}), 
and the physical contributions to the changes in vertical velocities in the dry decomposition (section \ref{sec:drydecomp}) 
and moist decomposition (section \ref{sec:moistdecomp}). 
We briefly describe the results for a second GCM and for daily precipitation extremes (section \ref{sec:gfdl_daily}) 
before giving our conclusions (section \ref{sec:conclusions}).

\section{Simulations and definition of extreme precipitation events}\label{sec:data}

We primarily focus on coupled model output from the Community Earth System
Model Large-Ensemble Project (CESM-LE) \citep{Kay2015} because of its high
resolution and availability of many ensemble members.  The contribution of
GFDL-CM3 \citep{Donner2011} to CMIP5 \citep{Taylor2012} is also analyzed as a
point of comparison. 
Climate change is defined as the difference between the
historical and RCP8.5 scenario simulations, and percentage changes of physical
quantities are reported normalized by the historical values and the change in
global-mean surface air temperature.  For CESM-LE, the data are on a
1.25$^{\circ}$ longitude by 0.94$^{\circ}$ latitude grid, and we use 1991-2000
for the historical climate and 2071-2080 for RCP8.5.  We are able to analyze
only 6 out of 40 ensemble members of CESM-LE because of storage constraints and
the cost of extracting data to solve the QG-$\omega$
equation for many events.  
The ensemble members used are 1 through 5 and 35 (member 35 was used
because it had previously been downloaded).  For GFDL-CM3, there is only one
ensemble member, the data are on a coarser 2.5$^{\circ}$ longitude by
2$^{\circ}$ latitude grid, and we use 1980-1999 for the historical simulation
and 2081-2100 for RCP8.5.  We focus on 6-hourly precipitation extremes in
CESM-LE, but we also describe results for GFDL-CM3 and daily precipitation
extremes (Section \ref{sec:gfdl_daily}).

6-hourly precipitation rates are directly available for CESM-LE and 
are averaged from 3-hourly precipitation rates for GFDL-CM3 such that 
the centers of the 6-hourly precipitation averaging periods correspond to the times of the instantaneous dynamical fields. 
The pressure vertical velocity ($\omega$) is not directly available and is
instead calculated from other fields
as described in Appendix A.
6-hourly instantaneous horizontal winds ($u$, $v$) and temperature ($T$) are
needed as inputs when solving the QG-$\omega$ equation, and these are linearly
interpolated in the vertical from a hybrid sigma coordinate to a pressure coordinate. 

\begin{figure*}
\centering
\noindent\includegraphics[width=0.8\linewidth]{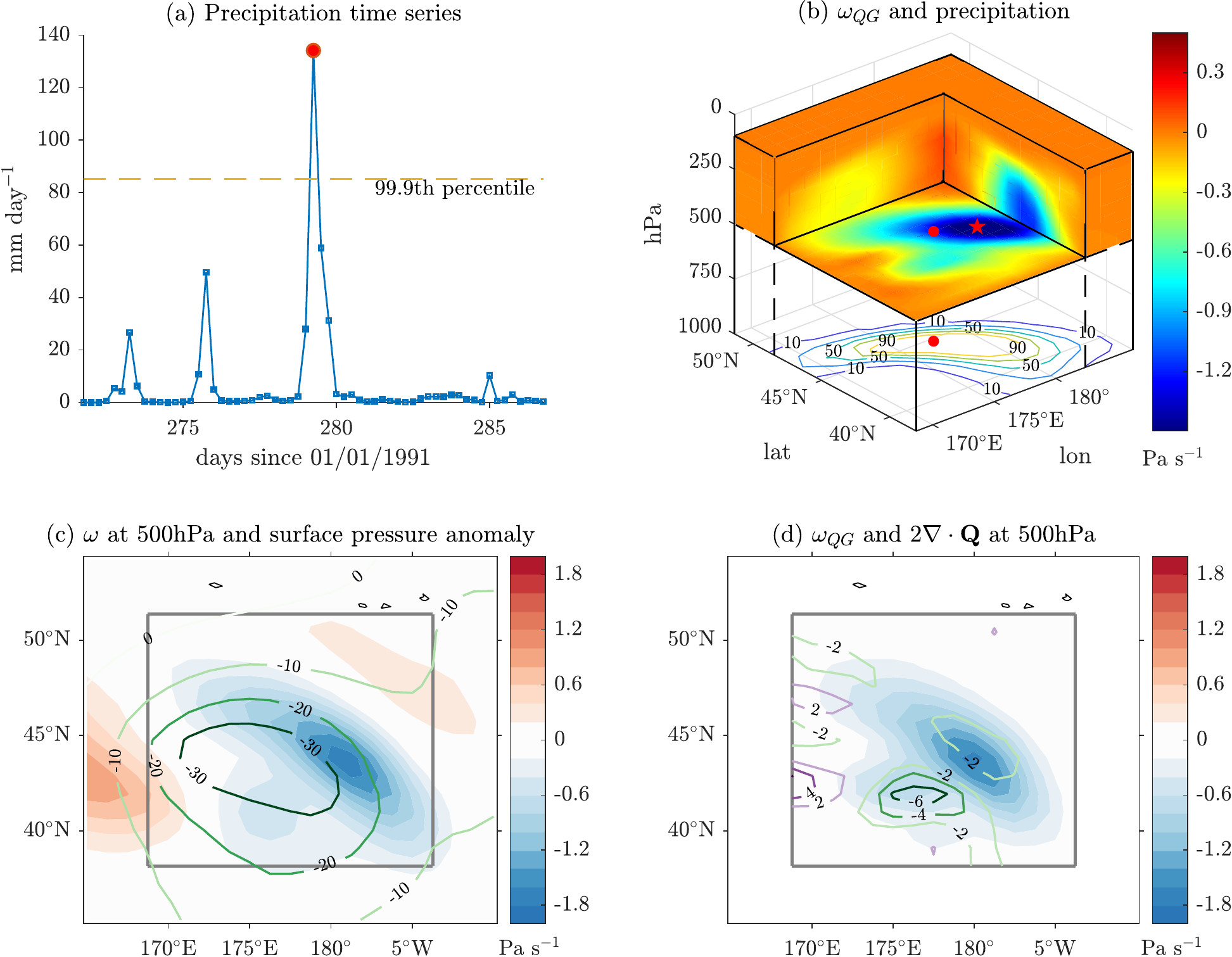}
\caption{ 
A typical extreme precipitation event from the historical climate in CESM-LE featuring strong upward motion in the center of the domain. (a) The extreme precipitation event (red dot) is defined as an exceedance of the 6-hourly precipitation rate at a given grid point (blue line with squares) relative to the $99.9^{\text{th}}$ percentile of the distribution at that grid point (yellow dashed line). (b) The precipitation rate for the event is shown by the contours at the bottom with an interval of 20 mm day$^{-1}$, and $\omega_{\text{QG}}$ is shown by the shading at 500hPa and above. The two red dots indicate the horizontal location of the extreme precipitation event at the surface, and the red star indicates the location at which we evaluate $\omega$ and $\omega_{\text{QG}}$ at 500hPa. (c) The surface pressure anomaly is shown by contours with an interval of 10hPa, and $\omega$ at 500hPa is shown in shading. 
(d) The divergence of the Q-vector field ($2\nabla\cdot\mathbf{Q}$) is shown with a contour interval of  $2\times 10^{-17}$ s$^{-3}$ Pa$^{-1}$ without zero contour, and  $\omega_{\text{QG}}$ at 500hPa is shown in shading.  Grey squares in (c) and (d) depict the domain used to invert the QG-$\omega$ equation for this event.}
\label{fig:example_event}
\end{figure*}

We define an extreme precipitation event at a grid point in a given climate as
a 6-hourly period over which the average precipitation rate exceeds its
99.9$^\text{th}$ percentile for that grid point and climate.  Instances with
zero precipitation are included when calculating percentiles as recommended by
\citet{Schar2016}, 
and thus the return period of these 6-hourly events is 250 days. 
Visual inspection of individual extratropical events
suggests that they are typically associated with precipitation structures in
extratropical cyclones rather than grid-point storms.  
A small fraction of events are excluded from the results due to issues such as numerical instability of the QG-$\omega$ inversion or the inverted vertical velocity at 500hPa being downward (see Appendix B for details of event selection). 
Taking the 6 ensemble members of
CESM-LE together, there are roughly 85 6-hourly extreme precipitation events to
be analyzed at each grid point in each climate. 
Using only one event when events are adjacent in time would have removed roughly 18\% of all events in the extratropics, 
but we chose not to remove them because that would change the weighting of events of 
different durations and thus complicate comparison to previous studies.

An example event that took place in the northern Pacific ocean in the modeled
fall season is shown in Fig.\;\ref{fig:example_event}.  The synoptic maps of this
event show a low pressure system with upward motion mostly on the eastern side
(shading in Fig.\;\ref{fig:example_event}c) that is partly driven by large-scale forcing as
represented by the Q-vector divergence (contours in Fig.\;\ref{fig:example_event}d).

\section{Numerical inversion of the QG-$\omega$ equation}\label{sec:qgomega}

To understand the behavior of the vertical velocities in the extreme
precipitation events, we numerically solve the QG-$\omega$ equation to give
$\omega_{\text{QG}}$ for all such events at grid points between $70^\circ$S and
$70^\circ$N.  For each event, we expand a three-dimensional domain centered
around the location of the event, as illustrated for our example event in
Fig.\;\ref{fig:example_event}b.  

For CESM-LE, the inversion domain extends 29 grid points ($\sim$3000km) 
in each horizontal direction and from 1000hPa to 100hPa in the vertical 
when there are no missing (subsurface) values in the domain.  
However, the domain can shrink to a minimum of 15 grid points simultaneously in both horizontal directions to avoid missing values where the surface pressure is below 1000hPa.  If this horizontal shrinking is not sufficient to avoid
missing values, the lower boundary is then moved up to levels as high as
550hPa.  For GFDL-CM3, the domain is chosen following the same approach except
that it varies between 15 grid points ($\sim$3300km) and 9 grid points as
necessary in each horizontal direction.  We apply these choices of domain size
to ensure that the domains have a minimum width of about 1500km. 
Changing the domain sizes to be smaller or bigger by 20\% was found to not affect the results.

We impose Dirichlet boundary conditions on all boundaries: $\omega_{\text{QG}}$
is set to climatological means on the lateral boundaries and to zero at both
the top and bottom boundaries. This bottom boundary condition is a
simplification that neglects topographic forcing and Ekman pumping, but the
results at 500hPa are nonetheless reasonably accurate, and the impact of
instead taking the exact boundary values from the GCM simulations is discussed
in Section \ref{sec:gfdl_daily}.

The QG-$\omega$ equation is written as
\begin{equation} 
\left(\nabla^2\sigma + f_0^2\frac{\partial^2}{\partial
p^2}\right)\omega_{\text{QG}} = Adv - \frac{\kappa}{p}\nabla^2J, 
\label{eq:QG_omega}
\end{equation}
where $f_0$ is
the Coriolis parameter evaluated at the center of the domain, $p$ is pressure,
$\kappa$ is the ratio of the gas constant to specific heat
capacity at constant pressure, and $J$ is the diabatic heating.  The static stability parameter $\sigma$ is given by $\sigma = -\frac{RT}{p\theta}\frac{\partial\theta}{\partial p}$ where $R$ is the gas constant for dry air, $T$ is temperature,  and $\theta$ is potential temperature. 
The advective forcing is given by
\begin{equation}
Adv = -2\nabla_h\cdot\mathbf{Q} + f_0\beta\frac{\partial v_g}{\partial p},
\label{eq:Adv}
\end{equation}
where $\beta$ is the meridional derivative of the Coriolis parameter and
$\nabla_h$ is the horizontal gradient. 
We calculate the Q-vector ($\mathbf{Q}$) in equation \eqref{eq:Adv} in spherical coordinates
\begin{equation}
    \begin{aligned}
    \mathbf{Q} 	& = -f_0\bigg[\frac{\partial u_g}{\partial p}\frac{1}{a\cos{\phi}}\bigg(\frac{\partial v_g}{\partial \lambda} + u_g\sin{\phi}\bigg) + \frac{\partial v_g}{\partial p}\frac{1}{a}\frac{\partial v_g}{\partial\phi}\bigg] \mathbf{i}+\\
			&-f_0\bigg[\frac{\partial u_g}{\partial p}\frac{1}{a\cos{\phi}}\bigg(-\frac{\partial u_g}{\partial \lambda} + v_g\sin{\phi}\bigg) - \frac{\partial v_g}{\partial p}\frac{1}{a}\frac{\partial u_g}{\partial\phi}\bigg] \mathbf{j}, 
    \end{aligned}
    \label{eq:Q_vector}
\end{equation}
where $a$ is Earth's radius,
$u_g,v_g$ are the zonal and meridional geostrophic winds, 
$\lambda,\phi$ are longitude and latitude, 
and $\mathbf{i},\mathbf{j}$ are the zonal and meridional unit vectors, respectively.  
This form of Q-vector is the same as that given by equation (19) in
\citet{Dostalek2017}, 
except that we approximate the Coriolis parameter with $f_0$ at the center of the domain of
each event.  
We choose to use the Q-vector form of the QG-$\omega$ equation \citep{Hoskins1978}
because it yields a smoother advective forcing ($Adv$) than the traditional
form, and this is likely because it avoids a cancellation between terms in the
traditional form that can lead to substantial errors when the derivatives are
approximated numerically.  To minimize the influence of gravity waves, the
geostrophic winds are calculated as the rotational component of the horizontal
wind \citep{Nielsen-Gammon2008} which is obtained through inverting the
relative vorticity on a global spherical grid in pressure coordinates.
We use smoothing of temperature and a spatially-varying static stability to improve the accuracy and stability of the numerical inversions as described in detail in Appendix A.

The QG-$\omega$ equation is inverted in each event domain in spherical coordinates
using a 3D variant \citep{Zedan1983,Ferziger2002} of the strongly implicit
method \citep{Stone1968}, similar to the approach of Shaevitz et al, 2016
(arXiv:1603.01317).  
The inverted field of $\omega_{\text{QG}}$ is shown for the example event as shadings in Fig.\;\ref{fig:example_event}b, d, and it agrees well
with the vertical velocity calculated from model output (shading in Fig.\;\ref{fig:example_event}c). 

In this paper we focus on $\omega_{\text{QG}}$ at 500hPa for simplicity. 
However, as can be seen from Fig.\;\ref{fig:example_event}b,
the maximum of surface precipitation and 
the maximum of $-\omega_{\text{QG}}$ at 500hPa need not be colocated in the horizontal. We choose to focus on the local maximum value of $-\omega_{\text{QG}}$ at 500hPa that is closest in horizontal distance to the extreme precipitation event. 
This local maximum is taken to represent the center of the dominant upward motion at 500hPa
which likely contributes substantially to the surface precipitation 
over the 6-hour (or daily) averaging period.
However, if the local maximum is too far away from the event, 
we discard the event according to our event exclusion criteria in Appendix B.
For consistency, the values of $\omega$ and other variables that we plot in all subsequent figures are taken from the same location as the local maximum of $-\omega_{\text{QG}}$ for each event.  Similarly, averages across extreme precipitation events for all variables are evaluated at these locations.

\begin{figure*}
\centering
\includegraphics[width=1.0\linewidth]{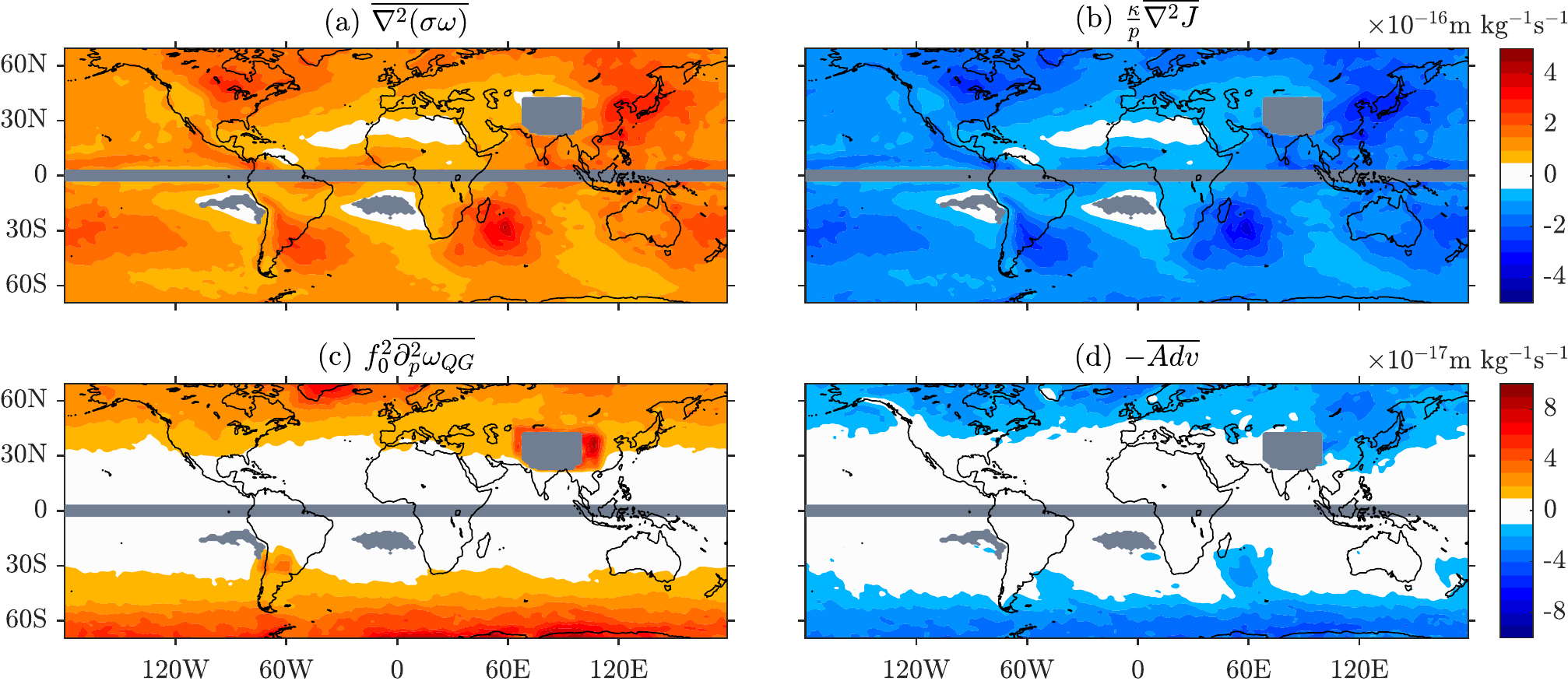}
\caption{
Event-mean of terms in the QG-$\omega$ equation (Eq. 1) at 500hPa for precipitation extremes in CESM-LE.  Terms on the right-hand side of the equation are shown with a minus sign so that the sum is zero. Shown are terms involving the
(a) static stability ($\overline{\nabla^2\sigma\omega_{\text{QG}}}$),
(b) diabatic heating ($\frac{\kappa}{p}\overline{\nabla^2 J}$),
(c) vertical curvature ($f_0^2\overline{\partial_p^2\omega_{\text{QG}}}$), and
(d) advective forcing ($-\overline{Adv}$).
Note that (a) and (b) have much larger magnitudes than (c) and (d) and are shown
with a different color bar.
Tropical regions between 3$^\circ$S and 3$^\circ$N are masked due to a larger 
fraction of unstable solutions in the inversion of the QG-$\omega$ equation.
Also masked are the Tibetan Plateau and grid points with fewer than 30 events. 
A 1-2-1 filter was applied 3 times in each direction to the results for clarity.
}
\label{fig:terms_2D}
\end{figure*}

Averaging across all events for each grid point, which we denote by an overbar, the static stability
term ($\overline{\nabla^2\sigma\omega_{\text{QG}}}$) and the diabatic heating
term ($-\frac{\kappa}{p}\overline{\nabla^2 J}$) are the dominant terms in the
QG-$\omega$ equation and are of similar magnitudes 
as shown in Fig.\;\ref{fig:terms_2D} (see
Fig.\;S1
for the zonal mean in the extratropics). 
This similarity is expected 
for events driven by large-scale 
dynamical forcing in the free troposphere since $\sigma+(J \kappa)/(p \omega)$ may then be viewed as a
measure of the moist stability which will be close to zero for a stratification
that is close to moist adiabatic in extreme precipitation events
\citep{OGorman2015}. The vertical curvature term 
($f_0^2\overline{\partial_p^2\omega_{\text{QG}}}$) and the advective
forcing term ($\overline{Adv}$) are of comparable magnitudes to each other 
but are considerably smaller in magnitude than the other two terms.

We have shown in this section that $\omega_{\text{QG}}$ captures the behavior of $\omega$ in one example event. Further validation of $\omega_{\text{QG}}$ and its response to climate change is given in the next section.

\section{Vertical velocities associated with precipitation extremes and their response to climate change}\label{sec:omega_vs_omega_QG}

\begin{figure}[ht]
\centering
  \includegraphics[width=1.0\linewidth]{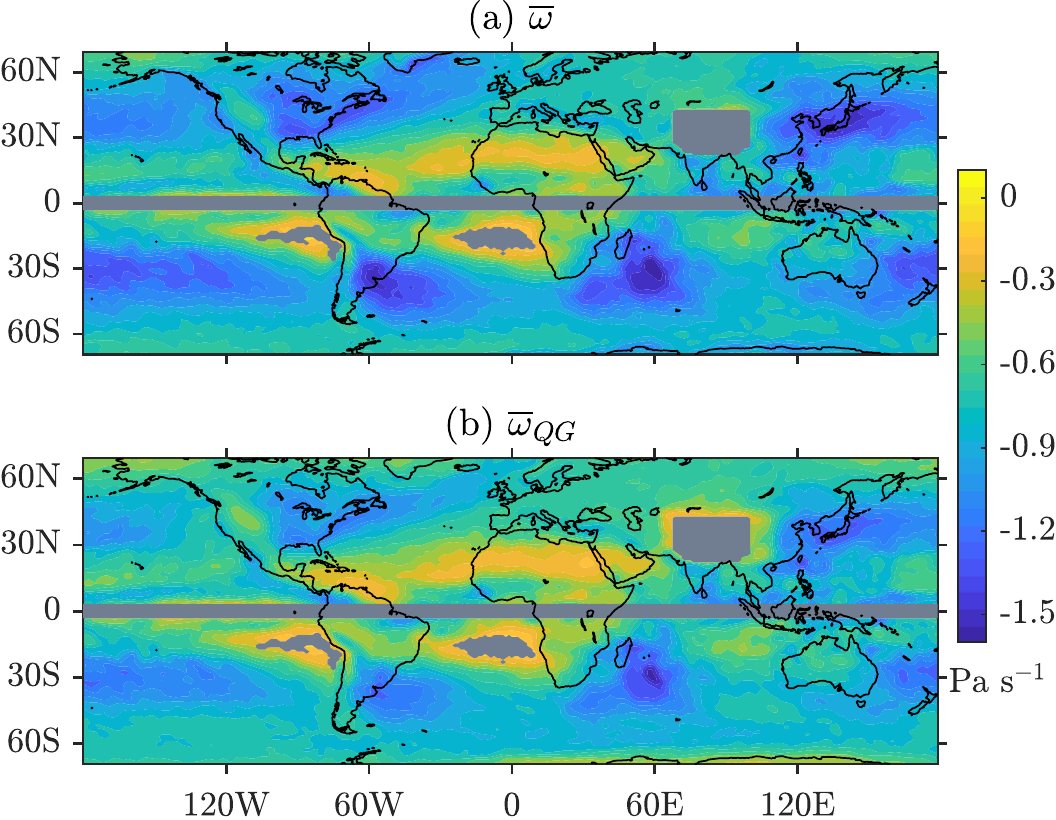}
\caption{ 
(a) $\overline{\omega}$ and (b) $\overline{\omega}_{\text{QG}}$ in Pa s$^{-1}$ at 500hPa 
associated with precipitation extremes in the historical simulations with CESM-LE. 
Masking and smoothing are as in Fig.\;\ref{fig:terms_2D}.
}
\label{fig:omega_vs_omega_QG}
\end{figure}

The instantaneous vertical velocity averaged over all extreme 
precipitation events at one grid point 
($\overline{\omega}$) is interpreted as the vertical velocity
associated with precipitation extremes for that location.  
As shown for the historical simulations in Fig.\;\ref{fig:omega_vs_omega_QG}a, $\overline{\omega}$ at 500hPa is generally negative in extreme precipitation events, consistent with upward motion, and maximizes in strength in regions such as the extratropical storm tracks.
There are events associated with downward motion at 500hPa in the subtropical ocean deserts,
and these events typically have ascent at lower levels. 
However, we neglect events with downward motion at 500hPa according to our exclusion criteria because the downward motion at 500hPa is not contributing positively to the precipitation.

The spatial pattern of $\overline{\omega}_{\text{QG}}$ at 500hPa in the
historical climate closely resembles that of $\overline{\omega}$ (compare
Fig.\;\ref{fig:omega_vs_omega_QG}a and Fig.\;\ref{fig:omega_vs_omega_QG}b),
although the magnitude is underestimated by roughly 14\%  if averaged over the
extratropics of both hemispheres between 30$^\circ$ and 70$^\circ$ in the historical and the RCP8.5 simulations. 
The underestimation mostly results from not using the exact lateral and bottom boundary conditions 
(see section \ref{sec:gfdl_daily}), 
and it is most pronounced around
the Tibetan Plateau, the Rocky Mountains, and Antarctica.

\begin{figure*}
\noindent\includegraphics[width=1.0\linewidth]{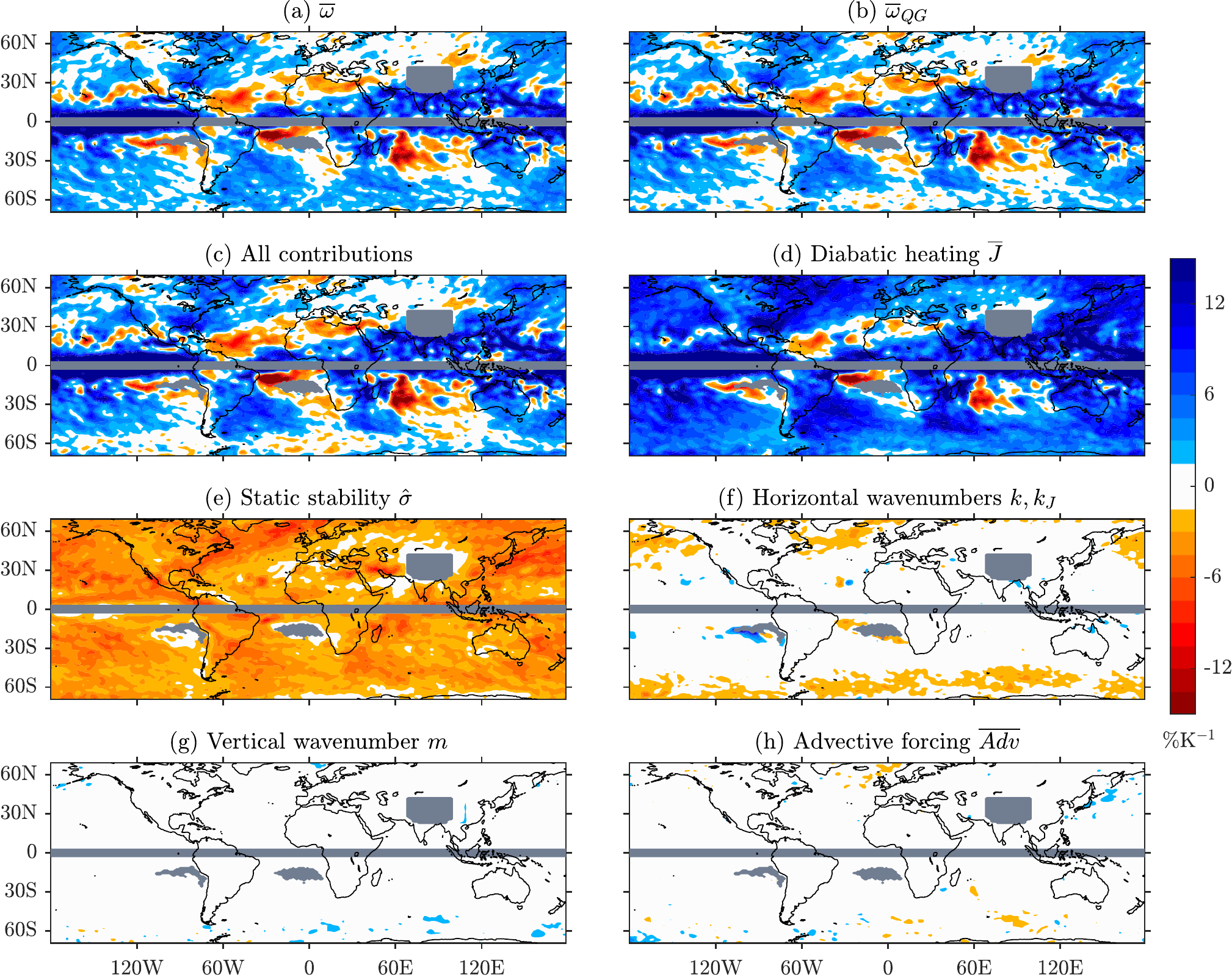}
\caption{ 
Percentage changes of vertical velocities at 500hPa 
associated with precipitation extremes in CESM-LE, 
and the contributions to these changes in the {\it dry} decomposition. 
Shown are changes in 
(a) $\overline{\omega}$ and
(b) $\overline{\omega}_{\text{QG}}$,
(c) the sum of all contributions, 
and contributions from changes in
(d) diabatic heating $\overline{J}$,  (e) static stability $\hat{\sigma}$, (f) horizontal wavenumbers $k$ and $k_J$, 
(g) vertical wavenumber $m$, and (h) advective forcing $\overline{Adv}$. 
The percentage changes are relative to the historical climate and are normalized by the increase in global-mean surface air temperature. 
Masking and smoothing are as in Fig.\;\ref{fig:terms_2D}. 
}
\label{fig:drydecomp}
\end{figure*}

\begin{figure*}
\noindent\includegraphics[width=1.0\linewidth]{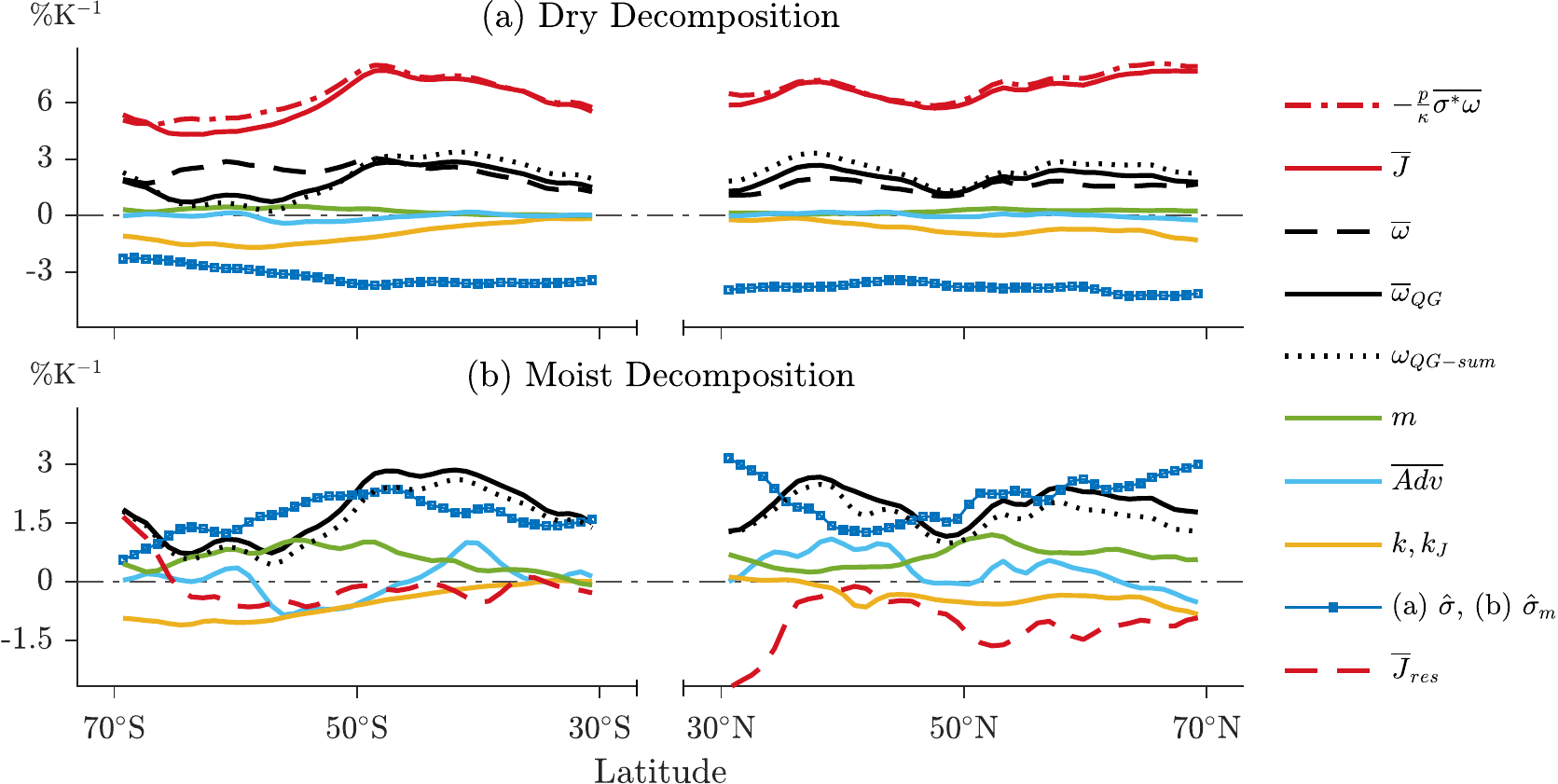}
\caption{ 
Percentage changes in zonal-mean $\overline{\omega}$ (black dashed) and $\overline{\omega}_{\text{QG}}$ (black solid) at 500hPa associated with extratropical precipitation extremes in CESM-LE, and the contributions in the (a) dry and (b) moist decompositions. 
Both panels show the sum of all contributions (black dotted) and the contributions from changes in vertical wavenumber (green), advective forcing (light blue), and the combined contribution of horizontal wavenumbers $k$ and $k_J$ (yellow).
Panel (a) also shows contributions from changes in diabatic heating (red solid), its approximation using Eq.\;(\ref{eq:J_latent}) (red dash-dotted), and dry static stability (solid blue with boxes). Panel (b) also shows contributions from changes in residual diabatic heating in Eq.\;\eqref{eq:residual_heating} (red dashed) and moist static stability (solid blue with boxes). 
The percentage changes are relative to the historical zonal-mean values and are normalized by the increase in global-mean surface air temperature. 
A 1-2-1 filter was applied once to each line for clarity.}
\label{fig:zonaldecomp}
\end{figure*}

The response of $\overline{\omega}$ to climate change
is a strengthening or little change in the extratropics,
with weakening primarily confined to parts of the
subtropical oceans and nearby land regions (Fig.\;\ref{fig:drydecomp}a). 
This is broadly consistent with the dynamical contribution                
calculated by \citet{Pfahl2017}
for projected changes in daily precipitation extremes, 
although for daily precipitation extremes there is a more 
mixed response in the extratropics (see Section \ref{sec:gfdl_daily}).
The zonal-mean response of $\overline{\omega}$ shows a strengthening at all
extratropical latitudes (Fig.\;\ref{fig:zonaldecomp}a), and the
extratropical-average response is relatively modest at 1.9\% K$^{-1}$. This 
extratropical-average response is calculated by taking the
percentage change of the zonal mean at each latitude, averaging between
30$^\circ$ and 70$^\circ$ latitude with area weighting in both hemispheres, and
normalizing by the increase in global-mean surface air temperature. 
Extratropical-average responses calculated in this way are
summarized in Tables \ref{tab:drydecomp} and \ref{tab:moistdecomp}. 

The response of $\overline{\omega}$ to climate change is well captured by $\overline{\omega}_{\text{QG}}$ in terms of the regional response
(compare Figs.\;\ref{fig:drydecomp}a and b) and in the zonal average (Fig.\;\ref{fig:zonaldecomp}a), and thus we will analyze
the contributions to changes
$\overline{\omega}_{\text{QG}}$ to better understand this response. 

\section{Dry Decomposition}\label{sec:drydecomp}

\begin{figure}[t]
\centering
    \centering
    \includegraphics[width=1.0\linewidth]{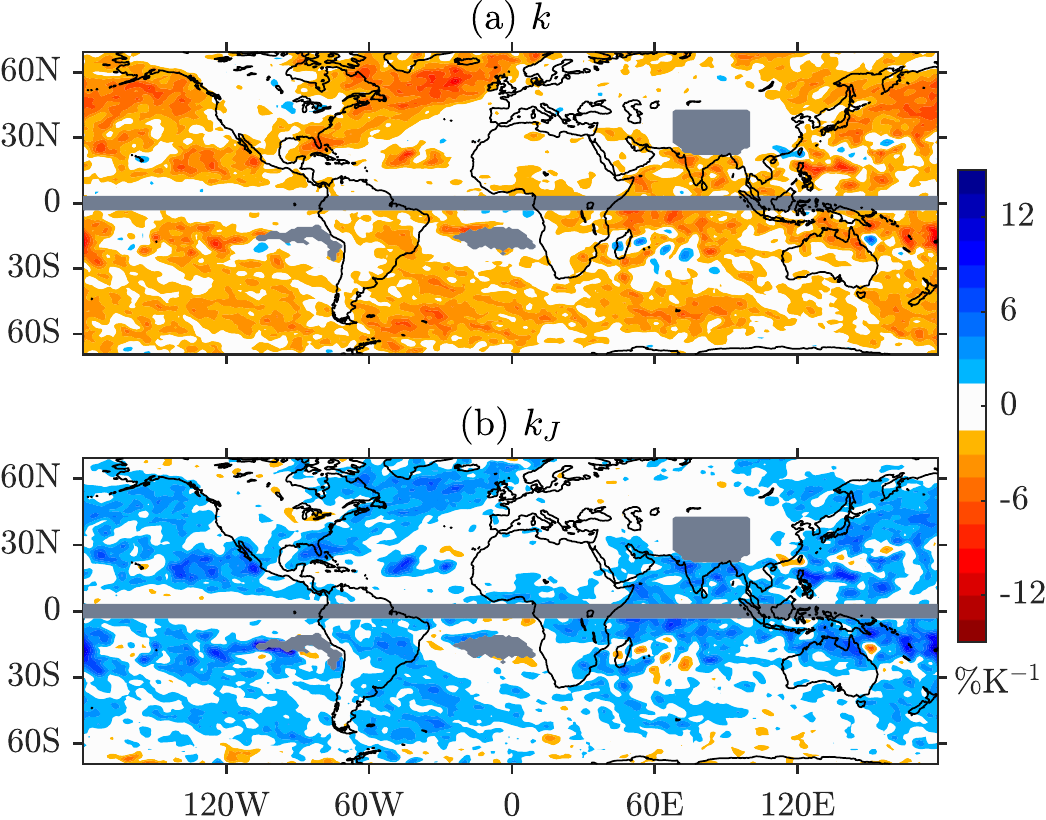}
\caption{Contributions in the dry decomposition from changes in (a) $k$ and (b) $k_J$ to the change in $\overline{\omega}_{\text{QG}}$ at 500hPa associated with 6-hourly precipitation extremes in CESM-LE. Percentage changes are shown relative to the historical climate and normalized by the change in global-mean surface air temperature. Masking and smoothing are as in Fig.\;\ref{fig:terms_2D}.
}
\label{fig:k2_vs_kJ2}
\end{figure}

We decompose changes in $\overline{\omega}_{\text{QG}}$ at 500hPa into different physical contributions according to the QG-$\omega$ equation. 
We begin with a \textit{dry decomposition} in which the diabatic heating ($J$), dominated by latent heating, is considered as an external forcing. 
For each event, we focus on the local maximum of $-\omega_{\text{QG}}$ at 500hPa as described at the end of section \ref{sec:qgomega}. 
The QG-$\omega$ equation (Eq.\;\ref{eq:QG_omega}) involves second derivatives in space, and 
thus we need a method to represent the length scales involved in these derivatives
in order to quantify their contribution to changes in $\omega_{\text{QG}}$. 
We do this using effective wavenumbers that are defined in terms of the second derivatives of the fields as
$\tilde{k}^2 = -\nabla^2(\sigma\omega_{\text{QG}})/(\sigma\omega_{\text{QG}})$, $\tilde{k}_J^2 = -\nabla^2J/J$, and $\tilde{m}^2 = -\partial^2_p\omega_{\text{QG}}/\omega_{\text{QG}}$. 
We refer to them as effective wavenumbers rather than wavenumbers because a given disturbance will have more than one wavelength. 
Note that $\tilde{k}^2$ accounts for the combined spatial structure of $\omega_{\text{QG}}$ and $\sigma$. 
Eq.\;\eqref{eq:QG_omega} then becomes
\begin{equation}
    -\tilde{k}^2\sigma\omega_{\text{QG}}  - f_0^2\tilde{m}^2 \omega_{\text{QG}} = Adv + \frac{\kappa}{p}\tilde{k}_J^2J,
    \label{eq:QG_omega_1}
\end{equation}
which implies that apart from contributions of $Adv$ and $J$, 
the magnitude of $\omega_{\text{QG}}$ is also affected by the local curvatures of $\sigma \omega_{\text{QG}}$ and $J$. 

To determine the average contributions to changes in $\omega_{\text{QG}}$ we need to average Eq.\;\eqref{eq:QG_omega_1} across events in a given climate, but a simple average across events will lead to averages of products of variables which are difficult to disentangle since the variables vary across events and are not independent.  We instead define a composite event that represents all events that occur at a given gridbox in a given climate and which satisfies the QG-$\omega$ equation.
We define composite effective wavenumbers as
$k^2 = -\overline{\nabla^2(\sigma\omega_{\text{QG}})}/\overline{\sigma\omega_{\text{QG}}}$, $k_J^2 = -\overline{\nabla^2J}/\overline{J}$, 
$m^2 = -\overline{\partial^2_p\omega_{\text{QG}}}/\overline{\omega}_{\text{QG}}$,
and a composite static stability as $\hat{\sigma} = \overline{\sigma\omega_{\text{QG}}}/\overline{\omega}_{\text{QG}}$. The event-average of Eq.\;\eqref{eq:QG_omega} at each gridbox can then be written exactly as
\begin{equation}
    -k^2\hat{\sigma}\overline{\omega}_{\text{QG}}  - f_0^2m^2 \overline{\omega}_{\text{QG}} = \overline{Adv} + \frac{\kappa}{p}k_J^2\overline{J}.
    \label{eq:QG_omega_2}
\end{equation}
Our focus on the local maximum of $-\omega_{\text{QG}}$ helps to ensure that $k^2$ and $k_J^2$ are positive.
For the rest of the paper, we will not refer to the wavenumbers of an individual event in Eq.\;\eqref{eq:QG_omega_1} 
but only their composite forms in Eq.\;\eqref{eq:QG_omega_2}. 
Although the composite static stability $\hat{\sigma}$
cannot be completely separated from $\omega_{\text{QG}}$ 
and is therefore affected by the vertical velocity, 
it nonetheless agrees with what would be obtained from a simple average 
over the extreme events ($\overline{\sigma}$). 
Our results for the dry decomposition would be similiar if we used $\overline{\sigma}$ instead of $\hat{\sigma}$, 
but the use of a composite static stability becomes important for maintaining
accuracy
in the moist decomposition in section \ref{sec:moistdecomp} in which the smaller moist static stability can have
large fractional variations across events. 

Equation \eqref{eq:QG_omega_2} can then be solved for $\overline{\omega}_{\text{QG}}$ as
\begin{equation}
    \overline{\omega}_{\text{QG}} = -\frac{\overline{Adv} + \frac{\kappa}{p} k_J^2 \overline{J}}{k^2 \hat{\sigma} + f_0^2m^2}. 
    \label{eq:drydecomp}
\end{equation}
We use a linear expansion (first-order Taylor expansion) of Eq.\;\eqref{eq:drydecomp} about the historical values to decompose the response of
$\overline{\omega}_{\text{QG}}$ to climate change into contributions from changes in
static stability ($\hat{\sigma}$), horizontal wavenumbers ($k$, $k_J$),
vertical wavenumber ($m$), QG forcing ($\overline{Adv}$) and diabatic heating
($\overline{J}$).  The effects of changes in the horizontal wavenumbers are 
combined because they offset one another as discussed below. The addition of
all the contributions approximately reconstructs the
change in $\overline{\omega}_{\text{QG}}$ (Fig.\;\ref{fig:drydecomp}c).

The largest contributions are from changes in diabatic heating (${J}$) 
and static stability ($\sigma$) (Fig.\;\ref{fig:drydecomp}d,e). 
We later show in section \ref{sec:moistdecomp} that $J$ is well-approximated by latent heating in saturated moist-adiabatic ascent, 
and thus $J$ depends on the vertical velocity and the amount of water vapor at saturation.  
The contribution of changes in $J$ is mostly a strengthening with an extratropical average contribution of 6.4\% K$^{-1}$, 
consistent with the increasing saturation vapor pressure.
But the contribution from changes in $J$ can also be negative where a sufficiently large weakening of upward motion 
cancels the strengthening from higher saturation vapor pressure.\footnote{The circular nature of this argument, which is we are trying to explain changes in vertical velocities through changes in $J$ and then explaining changes in $J$ through changes in vertical velocities, is part of the motivation to introduce a moist decomposition in the next section.\label{footnote1}}
The contribution of changes in ${\sigma}$ is almost uniformly a weakening, 
consistent with the projected increase of tropospheric dry static stability with warming \citep{Frierson2006}. 
In the extratropical average, the contribution from increases in static stability (-3.6\% K$^{-1}$) 
offsets a substantial portion of the effect of increased diabatic heating (6.4\% K$^{-1}$). 

Changes in both vertical wavenumber
and advective forcing contribute little in the dry decomposition, with
extratropical-average contributions of 0.2\% K$^{-1}$ and 0.0\% K$^{-1}$,
respectively (Fig.~\ref{fig:drydecomp}g,h).  
The vertical wavenumber decreases because of an upward stretching of the
vertical velocity profile,
consistent with an increase in the
depth of the upward motion in extreme events as the climate warms
\citep{Singh2012,Fildier2017}, and
 leading to a positive contribution to the upward motion at 500hPa. 
It may seem surprising that changes in advective forcing make a negligible contribution
given that advective forcing is likely to be ultimately driving many of the
events.  Percentage changes in the Q-vector may be expected to be of similar
magnitude to percentage changes in eddy kinetic energy which are of order 1 or
2\% K$^{-1}$ seasonally in climate-model projections \citep{OGorman2010}.
The smaller contribution from changes in advective forcing
results from the dominance of the diabatic heating term over
the advective forcing term on the right-hand-side of the QG-$\omega$ equation
(Fig.\;\ref{fig:terms_2D}). This smaller role for advective forcing, at least in the dry decomposition, is consistent with a greater role for changes in thermodynamic enviroments rather than changes in large-scale dynamical environments as hypothesized by \citet{Nie2018}. 
However, the effect of $Adv$ is greater in the moist decomposition (section \ref{sec:moistdecomp}) due to an increased sensitivity when a moist static stability is used. 

The combined changes in the horizontal wavenumbers weaken the upward velocity at higher latitudes (Fig.~\ref{fig:drydecomp}f), 
but averaged over the extratropics, the combined contribution is only -0.8\% K$^{-1}$.  Both $k$ and $k_J$ increase substantially with climate warming (not shown), which implies a decrease in the horizontal length scale of the vertical velocity field.  
This consistent decrease in length scale differs from the more 
mixed response of regional increases and decreases in 
the length scale of ascent (T18) or of precipitation \citep{Dwyer2017} found in 
previous studies of extreme precipitation events,  
likely because different models, measures of length scale, and extremeness of events were used. 
However, the contributions from the increases in $k$ and $k_J$ partially cancel each other in the QG-$\omega$ equation (Fig.\;\ref{fig:k2_vs_kJ2}). 

To see why these contributions from changes in $k$ and $k_J$ partially cancel each other, it is helpful to approximate Eq.\;\eqref{eq:drydecomp} by
neglecting $\overline{Adv}$ and $f_0^2m^2$ consistent with the balance of
terms in Fig.\;\ref{fig:terms_2D} to give
$\overline{\omega}_{\text{QG}} \simeq 
-(\kappa k_J^2 \overline{J})/(p k^2 \hat{\sigma} )$. For changes between climates
denoted by $\delta$, we then have that
\begin{equation}
\frac{\delta\overline{\omega}_{\text{QG}}}{\overline{\omega}_{\text{QG}}} 
\simeq 
\frac{\delta k_J^2}{k_J^2}
-\frac{\delta k^2}{k^2}
+\frac{\delta \overline{J}}{\overline{J}}
-\frac{\delta \hat{\sigma}}{\hat{\sigma}},
\label{eq:approx_drydecomp}
\end{equation}
which is a good approximation for the changes in 
$\overline{\omega}_{\text{QG}}$ 
(Fig.\;S2).
Because latent heating in extreme precipitation is closely tied to upward motion, we expect $k \simeq k_J$ and thus according to
Eq.\;\eqref{eq:approx_drydecomp} the contributions from changes in $k$ and $k_J$  should offset one another, consistent with Fig.\;\ref{fig:k2_vs_kJ2}.
However, at higher latitudes (poleward of $50^\circ$) the increase in $k$ is larger than the increase in $k_J$, possibly because 
advective forcing plays a bigger role at these latitudes which weakens the
link between spatial variations in diabatic heating and vertical motion.
The bigger increase in $k$ than $k_J$ at higher latitudes yields a net weakening contribution to upward motion at these latitudes (Fig.~\ref{fig:drydecomp}f). 
The contributions from changes in $k$ and $k_J$ also offset
one another in the subtropics, and we find that the overall contribution from
changes in horizontal scale is substantially smaller from what would be expected based on the scaling analysis of T18 for this region as discussed in Appendix C.

The zonal-mean dry decomposition for the extratropics in Fig.\;\ref{fig:zonaldecomp} shows
a dominant role of changes in diabatic heating and dry static stability, and that 
the strengthening effect of increased diabatic heating is partially offset by the weakening effect of increased dry stability. 
We will see in the next section that when this partial offsetting is taken into account by introducing a moist static stability,
other factors such as changes in the advective forcing and the increase in depth of the upward motion become more important.

\begin{table}
\centering
\ra{1.}
\begin{tabular}{lrrcrr}
\toprule[0.1em]
		& \multicolumn{2}{c}{6-hourly} 		&&  \multicolumn{2}{c}{daily}\\
\cmidrule{2-3} \cmidrule{5-6}
					&CESM			&GFDL 		&&CESM 		&GFDL\\
\midrule[0.07em]
$\overline{\omega}$        &   1.9 &   2.0 &&        0.9 &        0.6 \\[0.05cm]
$\overline{\omega}_{\text{QG}}$   &   2.0 &   2.0 &&        1.5 &        0.7 \\[0.05cm]
$\overline{J}$             &   6.4 &   7.4 &&        5.8 &        6.4 \\[0.05cm]
$\hat{\sigma}$             &  -3.6 &  -4.0 &&       -3.1 &       -3.5 \\[0.05cm]
$\overline{Adv}$           &  0.0 &  -0.4 &&       -0.1 &       -1.2 \\[0.05cm]
$k, k_J$                   &  -0.8 &  -1.2 &&       -0.9 &       -1.3 \\[0.05cm]
$m$                        &   0.2 &   0.5 &&        0.1 &        0.5 \\[0.05cm]
\bottomrule[0.1em]
\end{tabular}
\caption{Changes (\% K$^{-1}$) in vertical velocities and contributions in the dry decomposition averaged over the extratropics (30$^\circ$ to 70$^\circ$ in both hemispheres) for 6-hourly and daily precipitation extremes with CESM-LE and GFDL-CM3. All events are calculated at the 99.9$^\text{th}$ percentile except for the daily events in GFDL-CM3 which are at the 99.5th percentile.}
\label{tab:drydecomp}
\end{table}

\section{Moist decomposition}\label{sec:moistdecomp}

\begin{figure}[ht]
\centering
    \centering
    \includegraphics[width=0.8\linewidth]{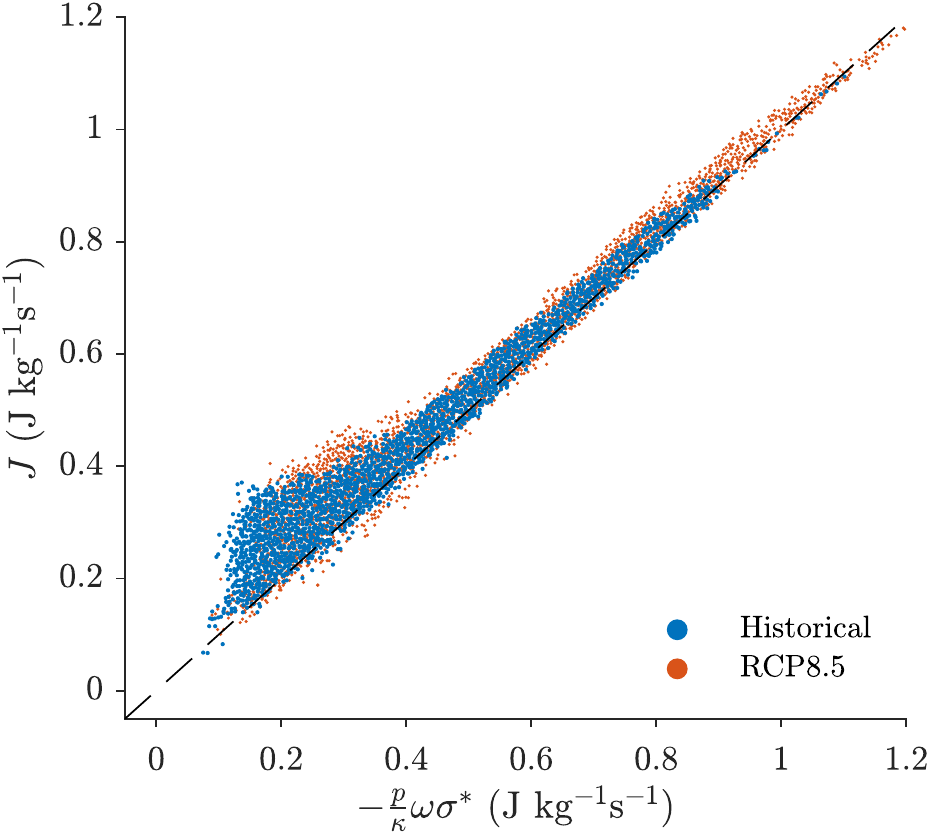}
\caption{Diabatic heating at 500hPa in 6-hourly extreme precipitation events versus its approximation by $-\frac{p}{\kappa}\omega\sigma^\ast$ in Eq.\;\eqref{eq:J_latent} based on saturated moist-adiabatic ascent. The dashed black line is the one-to-one line. Each dot is the mean of all the events at a grid point 
from the historical (blue) and RCP8.5 (orange) simulations with CESM-LE.
Results are shown for extratropical grid points between $30^\circ$ and $70^\circ$ for both hemispheres.
}
\label{fig:J_scatter}
\end{figure}

We introduce a moist decomposition of the QG-$\omega$ equation that links diabatic heating to the vertical velocity and thus avoids treating 
it as an external forcing.
The diabatic heating in extreme precipitation events is dominated by latent heating, and here we approximate it as the latent heating
associated with saturated moist-adiabatic ascent,
\begin{equation}
J = - \frac{p}{\kappa} \omega \sigma^\ast + \epsilon,
\label{eq:J_latent}
\end{equation}
where $\sigma^\ast$ is the static stability parameter for a moist-adiabatic lapse rate and $\epsilon$ is the error of the approximation.
Equation \eqref{eq:J_latent} follows from equation (1) of \citet{OGorman2011},
and similar parameterizations of condensational heating have been
used previously \citep{Emanuel1987}.  
For convectively-unstable events in which the stratification is close to
moist adiabatic, the approximation in Eq.\;\eqref{eq:J_latent} may also be
viewed as a simple quasi-equilibrium 
convective parameterization that maintains a 
moist-adiabatic vertical temperature profile 
when convection is forced by large-scale ascent.
The extreme precipitation 
events in our analysis are generally close to saturation, and the 
first term on the right hand side of
Eq.\;\eqref{eq:J_latent} is a good approximation for the diabatic
heating in these events, particularly for the strongest events as shown in Fig.\;\ref{fig:J_scatter}. 
This approximation also faithfully
captures the contribution of changes in diabatic heating to the changes in
$\overline{\omega}_{\text{QG}}$ in response to climate change (compare the red solid and dash-dotted lines in Fig.\;\ref{fig:zonaldecomp}a).  

Substituting Eq.\;\eqref{eq:J_latent} into the 
event-averaged QG-$\omega$ Eq.\;\eqref{eq:QG_omega_2} gives that 
\begin{equation}
    -k^2\hat{\sigma}\overline{\omega}_{\text{QG}}  
    + k_J^2 \overline{\sigma^\ast \omega} 
    - f_0^2m^2 \overline{\omega}_{\text{QG}} 
    = \overline{Adv} +
\frac{\kappa}{p}k_J^2\overline{\epsilon}.
\label{eq:diabatic_approx_1}
\end{equation}
The similarity of the first two terms on the left-hand side of Eq.\;\eqref{eq:diabatic_approx_1}
suggests that it will be useful to consider a moist static stability
that measures the deviation of the static stability ($\sigma$) from 
the static stability for a moist-adiabatic lapse rate ($\sigma^\ast$).
We therefore introduce a composite moist static stability
$\hat{\sigma}_{\text{m}} = \hat{\sigma}-\hat{\sigma}^\ast$ 
where $\hat{\sigma}^\ast = \overline{\omega_{\text{QG}}\sigma^\ast}/\overline{\omega}_{\text{QG}}$.
Rewriting Eq.\;\eqref{eq:diabatic_approx_1} using $\hat{\sigma}_{\text{m}}$
gives a moist QG-$\omega$ equation,
\begin{equation}
-k^2 \hat{\sigma}_{\text{m}} \overline{\omega}_{\text{QG}} 
- f_0^2m^2 \overline{\omega}_{\text{QG}} 
    = \overline{Adv} + \frac{\kappa}{p} k_J^2 \overline{J}_{\text{res}}, 
\label{eq:diabatic_approx_2}
\end{equation}
where we have defined the residual diabatic heating as
\begin{equation}
\overline{J}_{\text{res}} = -\frac{p}{\kappa} \left( \overline{\omega \sigma^\ast} - \frac{k^2}{k_J^2}\overline{\omega_{\text{QG}} \sigma^\ast} \right) + \overline{\epsilon}. 
\label{eq:residual_heating}
\end{equation}
$\overline{J}_{\text{res}}$ arises because not
all of the diabatic heating
is well-represented by the latent heating associated with
moist-adiabatic ascent at the velocity given by
 $\omega_{\text{QG}}$.
$\overline{J}_{\text{res}}$ is generally positive, and this is because 
the magnitude of $\omega$ is generally underestimated by $\omega_{\text{QG}}$. 
Differences between $k_J$ and $k$ and the error $\epsilon$ in Eq.\;\eqref{eq:J_latent} also contribute to $\overline{J}_{\text{res}}$.

Manipulating Eq.\;\eqref{eq:diabatic_approx_2} yields
\begin{equation}
        \overline{\omega}_{\text{QG}} = -\frac{\overline{Adv} + \frac{\kappa}{p}k_J^2\overline{J}_{\text{res}}}{k^2 \hat{\sigma}_{\text{m}}  + f_0^2 m^2}, 
    \label{eq:moistdecomp}
\end{equation}
Note that $k$ and $k_J$ in Eq.~\eqref{eq:moistdecomp} are the same as in the dry decomposition.
The moist static stability is smaller than the dry static stability and would be zero if the stratification in the extreme precipitation events was exactly moist-adiabatic.  A linear expansion of Eq.\;\eqref{eq:moistdecomp} about the historical
values gives the \textit{moist decomposition} of changes in $\overline{\omega}_{\text{QG}}$ into contributions
from changes in moist static stability ($\hat{\sigma}_m$), horizontal
wavenumbers ($k$, $k_J$), vertical wavenumber
($m$), advective forcing ($\overline{Adv}$) and residual diabatic
heating ($\overline{J}_{\text{res}}$).  

\begin{figure*}[ht]
\centering
    \centering
    \includegraphics[width=1.0\textwidth]{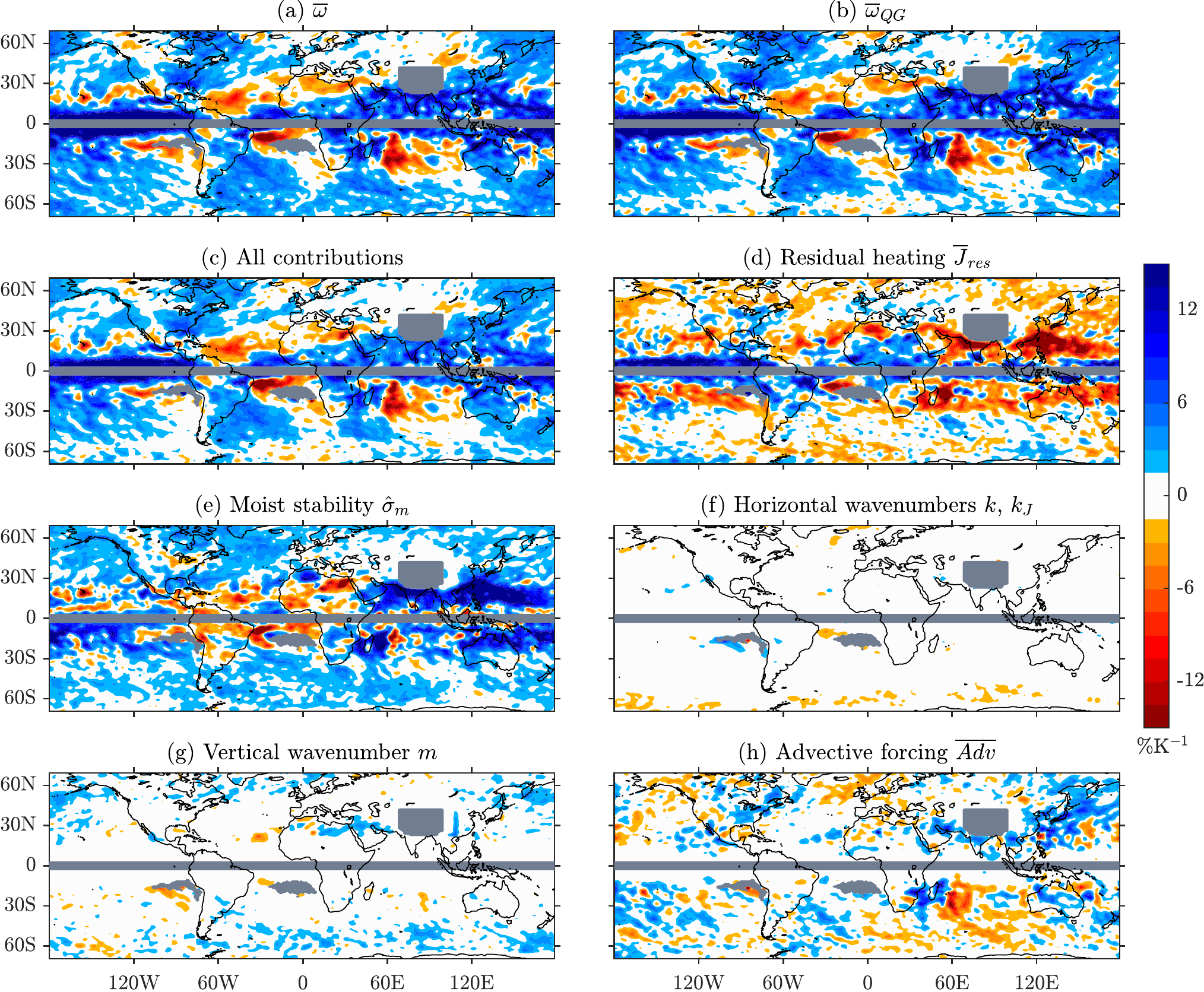}
\caption{
Percentage changes of vertical velocities at 500hPa 
associated with precipitation extremes in CESM-LE
and the contributions to these changes in the {\it moist} decomposition.
Shown are changes in
(a) $\overline{\omega}$ and
(b) $\overline{\omega}_{\text{QG}}$,
(c) the sum of all contributions, 
and contributions from changes in
(d) residual diabatic heating $\overline{J}_{\text{res}}$,  
(e) moist static stability $\hat{\sigma}_\text{m}$, 
(f) horizontal wavenumbers $k$ and $k_J$,
(g) vertical wavenumber $m$, and (h) advective forcing $\overline{Adv}$.
The percentage changes are relative to the historical climate and are normalized by the increase in global-mean surface air temperature.
Masking and smoothing are as in Fig.\;\ref{fig:terms_2D}.}
\label{fig:moistdecomp}
\end{figure*}

The moist decomposition is shown versus latitude and longitude in
Fig.\;\ref{fig:moistdecomp} and in the zonal mean for the extratropics in
Fig.\;\ref{fig:zonaldecomp}b.  The sum of all contributions
approximately reconstructs the total response (compare Fig.\;\ref{fig:moistdecomp}b and c). 
Remarkably, the contribution from changes in moist
static stability is similar in pattern and magnitude to the total response in
both the extratropics and subtropics (Fig.\;\ref{fig:moistdecomp}e).  
The contribution from changes in
residual diabatic heating has an extratropical average of only -0.6\% K$^{-1}$,
which is considerably smaller than the extratropical average of the contribution 
from diabatic heating in the dry decomposition (6.4\% K$^{-1}$).  
However, at lower latitudes the contribution from
changes in residiual diabatic heating is sizable 
and tends to partly offset
the contribution from changes in moist static stability 
(compare Figs.\;\ref{fig:moistdecomp}d and e).

Focusing on the extratropics, we find that decreases in the moist static
stability tend to strengthen the upward motion with an extratropical average
contribution of 1.9\% K$^{-1}$, in contrast to the weakening effect of
increases in dry stability in the dry decomposition. The decrease in moist
static stability corresponds to the extratropical stratification becoming
closer to moist adiabatic with warming, an effect that has also been found for
the mean extratropical stratification as the climate warms over a wide range in
an idealized GCM \citep[see Fig.\;9 in][]{OGorman2011}.  
Replacing the dry static stability with the smaller moist static stability and
the diabatic heating with the smaller residual diabatic heating increases the
importance of other terms in the moist decomposition compared to the dry
decomposition.  The contribution from decreases in vertical wavenumber ($m$) is
larger in the moist decomposition, with an extratropical-average contribution
of 0.6\% K$^{-1}$ as compared to 0.2\% K$^{-1}$ in the dry decomposition.
Similarly, the contribution from changes in advective forcing ($Adv$) is larger
in magnitude in the moist decomposition, with an extratropical-average
contribution of 0.2\% K$^{-1}$ as compared to 0.0\% K$^{-1}$ in the dry
decomposition.  However, $k^2$ is multiplied by the smaller moist static
stability and $k_J^2$ is multiplied by a smaller residual diabatic heating in
equation \eqref{eq:moistdecomp}, and the combined contribution from changes in
$k$ and $k_J$ in the moist decomposition is smaller in magnitude in the
extratropical average in the moist decomposition (-0.4\% K$^{-1}$) than in the
dry decomposition (-0.8\% K$^{-1}$).

Overall, the moist decomposition for CESM-LE suggests that increased upward
motion in extratropical precipitation extremes as the climate warms results primarily from decreased moist static
stability and increased depth of the upward motion. It is also noteworthy
that changes in moist static stability contribute much of the spatial pattern and magnitude of changes in vertical velocities in both the extratropics and subtropics, including some regions of weakening
of ascent in the subtropics, but we will see that residual diabatic heating
plays a greater role for GFDL-CM3 as discussed in the next section.

\begin{table}
\centering
\ra{1.}
\begin{tabular}{lrrcrr}
\toprule[0.1em]
	& \multicolumn{2}{c}{6-hourly} 		&&  \multicolumn{2}{c}{daily}\\
\cmidrule{2-3} \cmidrule{5-6}
					&CESM			&GFDL		&&CESM		&GFDL\\
\midrule[0.07em]
$\overline{\omega}$        &   1.9 &   2.0 &&        0.9 &        0.6 \\[0.05cm]
$\overline{\omega}_{\text{QG}}$   &   2.0 &   2.0 &&        1.5 &        0.7 \\[0.05cm]
$\overline{J}_{res}$       &  -0.6 &   0.9 &&       -0.8 &        1.4 \\[0.05cm]
$\hat{\sigma}_m$           &   1.9 &   1.2 &&        2.5 &        1.7 \\[0.05cm]
$\overline{Adv}$           &   0.2 &  -0.8 &&       -0.1 &       -2.5 \\[0.05cm]
$k, k_J$                   &  -0.4 &  -0.6 &&       -0.7 &       -0.7 \\[0.05cm]
$m$                        &   0.6 &   1.1 &&        0.3 &        0.9 \\[0.05cm]
\bottomrule[0.1em]
\end{tabular}
\caption{As in Table \ref{tab:drydecomp}, but for the moist decomposition.}
\label{tab:moistdecomp}
\end{table}

\section{Results for GFDL-CM3 and for daily precipitation extremes}\label{sec:gfdl_daily}

Changes in $\overline{\omega}$ are similar in magnitude for GFDL-CM3 as for
CESM-LE 
(Fig.\;S3)
However, because GFDL-CM3 has a coarser horizontal resolution, the horizontal
Laplacian terms in the QG-$\omega$ equation are smaller in magnitude, and
thus there is a relatively greater role for the vertical derivative and
advection terms as compared to CESM-LE 
(Fig.\;S4). 
As a result, the contributions from changes in vertical wavenumber and advective forcing are
of larger magnitude in the dry and moist decompositions for GFDL-CM3
(see 
Figs.\;S3, S5, S6
and Tables 1 and 2).
One other notable difference is that the moist decomposition for GFDL-CM3 
has larger contributions from changes in residual diabatic heating,
and changes in moist static stability contribute less of the overall 
pattern and magnitude of the change in vertical velocities.

The importance of changes in residual diabatic heating for the response in
GFDL-CM3 suggests that differences between $\omega$ and $\omega_{\text{QG}}$ as well
as diabatic heating not captured by the approximation for latent heating (the error $\epsilon$ in Eq.\;\ref{eq:J_latent}) are more
important for the response in this GCM.  
Differences between $\omega$ and $\omega_{\text{QG}}$ are caused by unbalanced
dynamics but also the 
boundary conditions that we use when inverting the QG-$\omega$ equation,
i.e., climatological means at the lateral boundaries and zero at the top and bottom.
We investigate the influence of the boundary conditions by performing
an additional set of inversions for GFDL-CM3 in which for each event, 
the bottom and lateral boundary values for the inversions are 
set to $\omega$ from the GCM output. This new setup leads to a 
500hPa-$\overline{\omega}_{\text{QG}}$ 
that more accurately reproduces 
$\overline{\omega}$, in that the underestimation of $\overline{\omega}$ by $\omega_{\text{QG}}$ is 8\%  
in the full-boundary case compared to 20\% in the default case.
However, the dry and moist decompositions remains broadly similar 
(compare Figs.\;S5 and S7 or Figs.\;S6 and S8)
which suggests that the boundary conditions
are not a key factor for our overall results. 

Daily extreme precipitation events are analyzed similarly to the 6-hourly
events with some modifications.  We calculate the 99.9$^\text{th}$-percentile daily
events for CESM-LE but the 99.5th-percentile daily events for GFDL-CM3 because
there are fewer events with daily precipitation compared to 6-hourly, and GFDL-CM3 has only one ensemble
member. With these choices, there are roughly 20 daily events per grid point
for CESM-LE and 30 for GFDL-CM3.
The precipitation rate for a given day is
calculated by averaging the four 6-hourly (interpolated) precipitation
rates for that day.  The static stability parameter ($\sigma$) is calculated
using the smoothed and time-averaged temperature over the day.  The vertical
velocity ($\omega$) shown in figures, 
advective forcing ($Adv$) and diabatic heating ($J$) are computed at each
6-hourly instance and then averaged to a daily value.

For daily precipitation extremes in CESM-LE 
(Figs. S9-S11),
the strengthening of upward motion in the extratropics is smaller in magnitude
(0.9\% K$^{-1}$  in the extratropical average) than for 6-hourly
precipitation extremes (1.9\% K$^{-1}$ ). For daily
precipitation extremes in GFDL-CM3 
(Figs.\;S12-S14),
the strengthening of upward motion is even less pronounced
(0.6\% K$^{-1}$ ) as compared to the 6-hourly extremes
(2.0\% K$^{-1}$ ).  The vertical velocity responses for daily
precipitation extremes in both GCMs have mixed positive and negative changes in
the extratropics 
(Figs.\;S9 and S12), 
consistent with the behavior of the dynamical
contribution to changes in daily precipitation extremes in the ensemble mean of
CMIP5 \citep{Pfahl2017}.  
In the extratropical average, the terms in the dry and
moist decompositions are of the same sign for daily extremes as for
6-hourly extremes, with the exception of the contribution
of changes in advective forcing for CESM-LE 
(see Tables \ref{tab:drydecomp} and \ref{tab:moistdecomp}). 

In the moist decomposition for both GCMs, 
more negative contributions from changes in advective forcing help to explain 
why the vertical velocities strengthen less for daily precipitation extremes 
as compared to 6-hourly precipitation extremes (Table 2).  
However, the weaker responses in vertical velocities at 500hPa at the daily
time scale does not translate to equivalently weaker changes in precipitation extremes at the daily time scale, 
perhaps due to our sole focus on 500hPa instead of the whole column, or differences in the thermodynamic response.
For example, in CESM-LE the extratropical-average response of precipitation extremes 
is 6.4\% K$^{-1}$ for 6-hourly events and 5.8\% K$^{-1}$ for daily events, 
which shows less of a difference than the responses in vertical velocities at 500hPa (1.9\% for 6-hourly events and 0.9\% for daily events).

\section{Conclusions}\label{sec:conclusions}

We have analyzed changes in vertical velocities associated with 6-hourly and
daily precipitation extremes at the 99.9$^{\text{th}}$ (or 99.5$^{\text{th}}$)
percentile in simulations of 21$^{\text{st}}$-century climate change with two
coupled GCMs.  For each extreme-precipitation event, we solved the QG-$\omega$
equation in a local domain, and the resulting vertical velocities at 500hPa
were shown to be in good agreement with the vertical velocity outputs from the
GCMs.  In general, upward motion in the extreme precipitation events is
strengthened in the extratropics in response to climate warming, and this was
first explained by a dry decomposition of the QG-$\omega$ equation in which
diabatic heating was treated as an external forcing.  According to the dry
decomposition, strengthening of upward motion by increased diabatic heating is
partly offset by increased dry static stability and, to a lesser extent,
changes in the horizontal extent of the extreme events.  Changes in horizontal
extent contribute little except at higher latitudes, and their small
contribution in the subtropics is in contrast to previous results based on a
scaling analysis of the QG-$\omega$ equation (T18).

The treatment of diabatic heating as an external forcing is a significant
limitation of the dry decomposition, especially when the overall aim is to
understand changes in surface precipitation rates which are directly related to
the column-integrated latent heating. Therefore, we derived a moist formulation
of the QG-$\omega$ equation in which diabatic heating is
approximated by the latent heating in moist-adiabatic saturated ascent.  
In the resulting moist decomposition for CESM-LE, much of the spatial pattern 
and magnitude of the change in vertical velocity is contributed by changes in moist static
stability, and this holds in both the extratropics and subtropics. 
In the zonal mean, the strengthening of
upwards velocities associated with extratropical precipitation extremes is
related to decreases in moist static stability and increases in the depth of
the upward motion under climate warming. 
In the coarser GFDL-CM3 GCM, however,
changes in residual diabatic heating, advective forcing and vertical extent 
play a greater role than in CESM-LE.

Our results suggest that the QG-$\omega$ equation is a useful diagnostic tool
to understand changes in vertical velocities in extreme precipitation events
in response to climate change.
The dry decomposition makes clear that factors such as increases
in dry static stability can at least partly offset the effect of increased
latent heating.
However, while the dry decomposition is self consistent, it does not take account of
the fact that latent heating is closely linked to $\omega_{\text{QG}}$.
The moist decomposition or similar moist dynamical approaches are then crucial to develop
a deeper understanding of the physical causes of changes in vertical velocities
in precipitation extremes, even if the smallness of the moist static
stability makes the moist decomposition challenging to perform accurately particularly at lower latitudes.
We have applied the QG-$\omega$ equation to
extreme precipitation events in global GCMs with relatively coarse
grid spacings and hydrostatic dynamics.
Since the QG-$\omega$ equation relies on both 
hydrostatic balance and quasi-geostrophic scaling, its application to
higher-resolution simulations or observational datasets
would require care in defining the balanced winds at appropriate length and time
scales \citep{Battalio2017}.

In future work, it would be interesting to further investigate the contribution
of changes in moist static stability to regional changes in extreme
precipitation. Residual diabatic heating, i.e., diabatic heating that is not
captured by moist-adiabatic saturated ascent driven by QG
dynamics, is also important and could be investigated using higher-order
equations for the vertical velocity \citep{Muraki1999,Davies2014} or by
including convective-scale dynamics as in the approach of \citet{Nie2018}. 
Conversely, one could also work to improve the approximation to diabatic heating (Eq.\;\ref{eq:J_latent})
so that the contribution of error ($\epsilon$) to residual diabatic heating is reduced. 
In addition, changes in advective forcing could be better understood by relating
them to changes in eddy kinetic energy \citep{OGorman2010} and changes in the
horizontal length scales of the geostrophic winds \citep{Kidston2010}.  It
would also be interesting to solve the QG-$\omega$ equation and analyze the dry
and moist decompositions for precipitation extremes in a wider range of GCMs
and in different seasons. 

%
\acknowledgments
We thank Matthieu Kohl, Ji Nie, and Neil Tandon for helpful discussions. 
We are also grateful to the three anonymous reviewers and the editor for their constructive comments. 
We acknowledge the CESM Large Ensemble Community Project and supercomputing resources provided by NSF/CISL/Yellowstone. 
We also acknowledge the World Climate Research Programme's Working Group on Coupled Modeling, which is responsible for CMIP, and we thank the NOAA Geophysical Fluid Dynamics Laboratory for producing and making available the output of GFDL-CM3.  
The datasets used in this paper are available at http://www.cesm.ucar.edu/projects/community-projects/LENS/data-sets.html and ftp://nomads.gfdl.noaa.gov/CMIP5. 
Data and code for the QG-$\omega$ inversion and figures are available at: https://github.com/dante831/QG-omega.git. 
This work was supported by NSF AGS 1552195 and the MIT Environmental Solutions Initiative.

\appendix[A]
\appendixtitle{Preprocessing of inputs to the QG-$\omega$ equation}
\label{appendix:preprocessing}

Since $\omega$ is not directly available as model output,
we calculate it from other fields following Eq.\;(3.11) of \citet{Simmons1981}: 
\begin{equation}
\frac{\kappa T\omega}{p} = - \frac{\kappa T}{p}\int_0^\eta\nabla\cdot\left(\mathbf{v}\frac{\partial p}{\partial\eta}\right)\mathrm{d}\eta + \kappa T\mathbf{v}\cdot\left(\frac{1}{p}\nabla p\right), 
\label{eq:continuity}
\end{equation}
where $\eta$ is the hybrid sigma vertical coordinate and other variables take their usual meaning.  We use discretized forms of the terms on the right-hand side of Eq.\;\eqref{eq:continuity} following Eqs.~(3.12) and (3.13) in \citet{Simmons1981}. 

To reduce numerical noise, the temperature ($T$) at each level is
smoothed by a 3-by-3 running-mean filter. 
The anomaly of the smoothed
temperature field from its horizontal mean over the event domain at each level is
rescaled so that it preserves the second moment of the unsmoothed field. 
We use this rescaling mainly to maintain the meridional gradient of the
temperature field.  

With the smoothed $T$ and the $\omega$ as calculated above, we compute the diabatic heating $J$ from
the thermodynamic equation 
under the hydrostatic approximation in pressure coordinates: 
\begin{equation}
J = c_p\left(\frac{\partial T}{\partial t} + \mathbf{v}\cdot\nabla T - \frac{p}{R}\omega\sigma\right), 
\label{eq:thermodynamic}
\end{equation}
where $\mathbf{v}$ is the horizontal wind vector and $c_p$ is the specific heat capacity.
Note that in order to calculate $J$ as accurately as possible, 
we don't apply quasi-geostrophic approximations when evaluating the
thermodynamic equation.

We calculate the static stability ($\sigma$) locally using the smoothed $T$
field, which means that we allow $\sigma$ to vary in the horizontal in order to
increase the accuracy of the inversion of the QG-$\omega$ equation.  This
approach does not compromise the derivation of the QG-$\omega$ equation as
long as $\sigma$ is kept inside the Laplacian operator.  
Compared to the traditional use of a homogeneous static stability calculated from the horizontal-mean
temperature over the domain ($\sigma_0$), 
the spatially-varying $\sigma$ also helps to prevent common situations 
where $\sigma_0$ is smaller than the local $\sigma$ in the center of the event, 
which is problematic for our moist decomposition (section \ref{sec:moistdecomp}) 
if it results in a negative moist static stability. 
However, horizontal variations in $\sigma$ can decrease the stability of
numerical inversions, especially for locations where $\sigma$ is close to zero,
promoting large local values and variations of $\omega_{\text{QG}}$.  
To minimize this instability, we set the spatially-varying $\sigma$ to 20\%
of $\sigma_0$ whenever the spatially-varying $\sigma$ falls below this value.
The resulting $\sigma$ field is also smoothed and rescaled according to the
same procedure as for the temperature that we described above. 
The rescaling of $\sigma$ helps to preserve the local variability of $\sigma$
around its spatial average, particularly near the maximum of upward motion. 
Not smoothing $T$ or $\sigma$ gives generally similar results except that 
more events are excluded and the changes in $\omega_{\text{QG}}$ are underestimated 
by roughly 25\%, although similar conclusions are reached from the dry decomposition.

\appendix[B]
\appendixtitle{Exclusion of events from the analysis}
\label{appendix:exclusion}

A small fraction of extreme precipitation events are excluded from all of our results. Events are excluded if any of the following conditions holds:  
\begin{enumerate}
\item The climatological mean surface pressure is lower than 550hPa.
\item The domain for the QG-$\omega$ inversion still includes grid points below the surface even when it is shrunk to the smallest allowed size as described in section \ref{sec:qgomega}. 
\item The closest local maximum of $-\omega_{\text{QG}}$ at 500hPa is more than 3 grid points away in CESM-LE (2 grid points in GFDL-CM3) from the horizontal location of the extreme precipitation event in either the zonal or meridional direction. 
\item More than $10\%$ of the grid points have negative $\sigma$ in the 3-dimensional inversion domain. 
\item The numerical inversion of the QG-$\omega$ equation is erroneous as manifested by NaN's due to numerical instability or unphysically large $\omega_{\text{QG}}$ (column-averaged absolute value larger than 10Pa s$^{-1}$ at the closest local maximum of $-\omega_{\text{QG}}$ at 500hPa).
\item The closest local maximum of $-\omega_{\text{QG}}$ at 500hPa is positive, since this implies downward motion at 500hPa which is not contributing to the extreme precipitation at the surface.
\end{enumerate}
Fewer than $8\%$ of the total events in the extratropics (between 30$^\circ$ and 70$^\circ$ latitude in both hemispheres) are excluded in a given climate and GCM. Therefore, the exclusion of these events is not expected to strongly affect our results.

\appendix[C]
\appendixtitle{Comparison with T18}
\label{appendix:compare_T18}

The results presented here for the contribution from changes in horizontal wavenumbers differs from the results of T18 as we now discuss. 
The dry decomposition in Fig.\;\ref{fig:drydecomp} may be directly compared with Fig.\;2 of T18 over latitudes 40$^\circ$S to 3$^\circ$S and 3$^\circ$N to 40$^\circ$N.  Over this common range of latitudes, our Fig.\;\ref{fig:drydecomp}f shows a very small contribution from changes in wavenumber (mostly below the contour interval of 1.5\% K$^{-1}$).  By contrast, Fig.\;2b of T18 shows widespread contributions of changes in horizontal length scale of much larger magnitude as interpreted by T18 through their scaling equation (3).  
T18 used daily rather than 6-hourly precipitation extremes, but our results are still different from T18 for daily extremes 
(Fig.\;S9f). 

T18 motivated their scaling equation (3) using a simplified balance
of the QG-$\omega$ equation that neglected the
static stability term, $\nabla^2(\sigma \omega_{\text{QG}})$, but this
is a dominant term at all latitudes when 
the QG-$\omega$ equation is numerically inverted for the extreme events analyzed here (see Fig.\;{\ref{fig:terms_2D}}).
T18 may have underestimated the
static stability term in the QG-$\omega$ equation because they scaled the Laplacian operator as
$\nabla^2 \sim -1/L^2$ with the horizontal length scale $L$ defined by $L^2 =
L_x^2 + L_y^2$ where $L_x$ and $L_y$ are the e-folding distances in x
and y directions. Instead, the Laplacian should typically be scaled
as $\nabla^2=\partial_x^2 + \partial_y^2 \sim -(1/L_x^2 + 1/L_y^2)$ 
(e.g., for the idealized case of a sinusoidal disturbance)
which is larger by a factor of $4$ when $L_x=L_y$. 
T18 also assumed that the length scale of horizontal variations in $\omega$ 
was similar to the Rossby deformation radius which could contribute to an
underestimate of the static stability term, and it is possible that the
assumption of a parabolic shape of the vertical structure of $\omega$ also
played a role.

A consequence of neglecting $\nabla^2(\sigma \omega_{\text{QG}})$ is
that changes in horizontal length scale associated with the vertical velocity are neglected whereas changes in horizontal length scale associated with diabatic heating are taken into account.  In the analysis presented here, these horizontal length scales enter through the composite effective wavenumbers, $k$ and $k_J$, and the
contributions of changes in $k^2$ and $k_J^2$ partially cancel each other, although less so at high latitudes (Fig.\;\ref{fig:k2_vs_kJ2}). This partial
cancellation occurs because $k_J\simeq k$, and the wavenumbers enter the approximate dry decomposition (Eq.\;\ref{eq:approx_drydecomp}) in the combination $\frac{\delta
k_J^2}{k_J^2} -\frac{\delta k^2}{k^2}$.  The approximate
dry decomposition given by Eq.\;\eqref{eq:approx_drydecomp} is
equivalent to the scaling Eq.\;(3) of T18 if the effects of
changes in both $k$ and dry static stability are neglected.  
(Eq.\;\ref{eq:approx_drydecomp}
is actually more similar to the scaling Eq.\;4 of T18 which has no
dependence on horizontal length scale but does include a dependence on static
stability.)
T18 used their Eq.\;(3) to argue for an important dependence on
horizontal length scale, and while this equation accounts for the
contribution of $k_J^2$, it doesn't account for $k^2$ which largely
offset the contribution of $k_J^2$ in the subtropics in our analysis.

T18 also found a match in spatial pattern between changes in vertical velocities and changes in horizontal length scale.
If we calculate the horizontal e-folding length scale $L$ as in T18, its changes show
a spatial pattern match with changes in vertical velocity for CESM-LE, but the spatial pattern match is much less apparent if GFDL-CM3 is considered or if $k$ is used instead of $L$. 
Thus the pattern match seems to have sensitivity to the GCM used and how the length scale is measured.
Note that in addition to using different GCMs than T18, we also analyze
less extreme events, i.e., return period of 250 days compared to 10-year maximum in T18.
Another difference is that we have to mask out parts of the subtropical ocean dry zones in the Southern Hemisphere because many of the precipitation extremes in this region are not associated with ascent at 500hPa, 
whereas the subtropical dry zones are a primary focus of T18. 
Nonetheless, we don't mask out the subtropical dry zones in the Northern
Hemisphere, and our inversions of the QG-$\omega$ equation and the approximate dry decomposition Eq.\;\eqref{eq:approx_drydecomp} suggest that changes in $k$ and $k_J$ will partly offset each other in the QG-$\omega$ equation, 
and thus the contribution from changes in horizontal length scales will yield a smaller contribution than found in T18, 
regardless of whether there's a pattern match between changes in length scale and $\omega$. 
To the extent that there is a pattern match between increases in horizontal length scale and weakening of upward motion in extreme precipitation events in certain GCMs and for certain measures of length scale as found by T18, it may require a different physical explanation than the QG-$\omega$ equation, which could be an interesting avenue for future work.

%






%
%
%

\bibliographystyle{ametsoc2014}
\bibliography{QG_omega_extremes}

\begin{thebibliography}{31}
\providecommand{\natexlab}[1]{#1}
\providecommand{\url}[1]{\texttt{#1}}
\renewcommand{\UrlFont}{\rmfamily}
\providecommand{\urlprefix}{URL }
\expandafter\ifx\csname urlstyle\endcsname\relax
  \providecommand{\doi}[1]{doi:\discretionary{}{}{}#1}\else
  \providecommand{\doi}{doi:\discretionary{}{}{}\begingroup
  \urlstyle{rm}\Url}\fi
\providecommand{\eprint}[2][]{\url{#2}}

\bibitem[{Battalio and Dyer(2017)Battalio, and Dyer}]{Battalio2017}
Battalio, M., and J.~Dyer, 2017: {The minimum length scale for evaluating QG
  omega using high-resolution model data}. \textit{Monthly Weather Review},
  \textbf{145}, 1659--1678, \doi{10.1175/mwr-d-16-0241.1}.

\bibitem[{Davies(2015)}]{Davies2014}
Davies, H.~C., 2015: {The quasigeostrophic omega equation: reappraisal,
  refinements, and relevance}. \textit{Monthly Weather Review}, \textbf{143},
  3--25, \doi{10.1175/MWR-D-14-00098.1}.

\bibitem[{Donner et~al.(2011)}]{Donner2011}
Donner, L.~J., and Coauthors, 2011: {The dynamical core, physical
  parameterizations, and basic simulation characteristics of the atmospheric
  component AM3 of the GFDL global coupled model CM3}. \textit{Journal of
  Climate}, \textbf{24}, 3484--3519, \doi{10.1175/2011JCLI3955.1}.

\bibitem[{Dostalek et~al.(2017)Dostalek, Schubert,, and DeMaria}]{Dostalek2017}
Dostalek, J.~F., W.~H. Schubert, and M.~DeMaria, 2017: {Derivation and solution
  of the omega equation associated with a balance theory on the sphere}.
  \textit{Journal of Advances in Modeling Earth Systems}, \textbf{9},
  3045--3068, \doi{10.1002/2017MS000992}.

\bibitem[{Dwyer and O'Gorman(2017)Dwyer, and O'Gorman}]{Dwyer2017}
Dwyer, J.~G., and P.~A. O'Gorman, 2017: {Changing duration and spatial extent
  of midlatitude precipitation extremes across different climates}.
  \textit{Geophysical Research Letters}, \textbf{44}, 5863--5871,
  \doi{10.1002/2017GL072855}.

\bibitem[{Emanuel et~al.(1987)Emanuel, Fantini,, and Thorpe}]{Emanuel1987}
Emanuel, K.~A., M.~Fantini, and A.~J. Thorpe, 1987: {Baroclinic instability in
  an environment of small stability to slantwise moist convection. Part I:
  two-dimensional models}. \textit{Journal of the Atmospheric Sciences},
  \textbf{44}, 1559--1573,
  \doi{10.1175/1520-0469(1987)044<1559:BIIAEO>2.0.CO;2}.

\bibitem[{Emanuel et~al.(1994)Emanuel, Neelin,, and Bretherton}]{Emanuel1994}
Emanuel, K.~A., J.~D. Neelin, and C.~S. Bretherton, 1994: {On large‐scale
  circulations in convecting atmospheres}. \textit{Quarterly Journal of the
  Royal Meteorological Society}, \textbf{120}, 1111--1143,
  \doi{10.1002/qj.49712051902}.

\bibitem[{Emori and Brown(2005)Emori, and Brown}]{Emori2005}
Emori, S., and S.~J. Brown, 2005: {Dynamic and thermodynamic changes in mean
  and extreme precipitation under changed climate}. \textit{Geophysical
  Research Letters}, \textbf{32}, L17\,706, \doi{10.1029/2005GL023272}.

\bibitem[{Ferziger and Peri{\'{c}}(2002)Ferziger, and
  Peri{\'{c}}}]{Ferziger2002}
Ferziger, J.~H., and M.~Peri{\'{c}}, 2002: \textit{{Computational methods for
  fluid dynamics}}. 3rd ed., Springer-Verlag, Berlin, Heidelberg, XIV, 426 pp.,
  \doi{10.1007/978-3-642-56026-2}.

\bibitem[{Fildier et~al.(2017)Fildier, Parishani,, and Collins}]{Fildier2017}
Fildier, B., H.~Parishani, and W.~D. Collins, 2017: {Simultaneous
  characterization of mesoscale and convective‐scale tropical rainfall
  extremes and their dynamical and thermodynamic modes of change}.
  \textit{Journal of Advances in Modeling Earth Systems}, \textbf{9},
  2103--2119, \doi{10.1002/2017MS001033}.

\bibitem[{Frierson(2006)}]{Frierson2006}
Frierson, D. M.~W., 2006: {Robust increases in midlatitude static stability in
  simulations of global warming}. \textit{Geophysical Research Letters},
  \textbf{33}, L24\,816, \doi{10.1029/2006GL027504}.

\bibitem[{Hoskins et~al.(1978)Hoskins, Draghici,, and Davies}]{Hoskins1978}
Hoskins, B.~J., I.~Draghici, and H.~C. Davies, 1978: {A new look at the
  $\omega$‐equation}. \textit{Quarterly Journal of the Royal Meteorological
  Society}, \textbf{104}, 31--38, \doi{10.1002/qj.49710443903}.

\bibitem[{Kay et~al.(2015)}]{Kay2015}
Kay, J.~E., and Coauthors, 2015: {The community earth system model (CESM) large
  ensemble project: A community resource for studying climate change in the
  presence of internal climate variability}. \textit{Bulletin of the American
  Meteorological Society}, \textbf{96}, 1333--1349,
  \doi{10.1175/BAMS-D-13-00255.1}.

\bibitem[{Kidston et~al.(2010)Kidston, Dean, Renwick,, and
  Vallis}]{Kidston2010}
Kidston, J., S.~M. Dean, J.~A. Renwick, and G.~K. Vallis, 2010: {A robust
  increase in the eddy length scale in the simulation of future climates}.
  \textit{Geophysical Research Letters}, \textbf{37}, L03\,806,
  \doi{10.1029/2009GL041615}.

\bibitem[{Muraki et~al.(1999)Muraki, Snyder,, and Rotunno}]{Muraki1999}
Muraki, D.~J., C.~Snyder, and R.~Rotunno, 1999: {The next-order corrections to
  quasigeostrophic theory}. \textit{Journal of the Atmospheric Sciences},
  \textbf{56}, 1547--1560,
  \doi{10.1175/1520-0469(1999)056<1547:tnoctq>2.0.co;2}.

\bibitem[{Nie and Sobel(2016)Nie, and Sobel}]{Nie2015}
Nie, J., and A.~H. Sobel, 2016: {Modeling the interaction between
  quasigeostrophic vertical motion and convection in a single column}.
  \textit{Journal of the Atmospheric Sciences}, \textbf{73}, 1101--1117,
  \doi{10.1175/JAS-D-15-0205.1}.

\bibitem[{Nie et~al.(2018)Nie, Sobel, Shaevitz,, and Wang}]{Nie2018}
Nie, J., A.~H. Sobel, D.~A. Shaevitz, and S.~Wang, 2018: {Dynamic amplification
  of extreme precipitation sensitivity}. \textit{Proceedings of the National
  Academy of Sciences}, \textbf{115}, 9467--9472,
  \doi{10.1073/pnas.1800357115}.

\bibitem[{Nielsen-Gammon and Gold(2008)Nielsen-Gammon, and
  Gold}]{Nielsen-Gammon2008}
Nielsen-Gammon, J.~W., and D.~A. Gold, 2008: {Dynamical diagnosis: a comparison
  of quasigeostrophy and Ertel potential vorticity}. \textit{Synoptic-Dynamic
  Meteorology and Weather Analysis and Forecasting. Meteorological Monographs},
  L.~Bosart, and H.~Bluestein, Eds., Vol.~33, American Meteorological Society,
  Boston, MA, chap.~9, 183--202, \doi{10.1007/978-0-933876-68-2_9}.

\bibitem[{O'Gorman(2010)}]{OGorman2010}
O'Gorman, P.~A., 2010: {Understanding the varied response of the extratropical
  storm tracks to climate change}. \textit{Proceedings of the National Academy
  of Sciences}, \textbf{107}, 19\,176--19\,180, \doi{10.1073/pnas.1011547107}.

\bibitem[{O'Gorman(2011)}]{OGorman2011}
O'Gorman, P.~A., 2011: {The effective static stability experienced by eddies in
  a moist atmosphere}. \textit{Journal of the Atmospheric Sciences},
  \textbf{68}, 75--90, \doi{10.1175/2010JAS3537.1}.

\bibitem[{O'Gorman(2015)}]{OGorman2015}
O'Gorman, P.~A., 2015: {Precipitation extremes under climate change}.
  \textit{Current Climate Change Reports}, \textbf{1}, 49--59,
  \doi{10.1007/s40641-015-0009-3}, \eprint{1503.07557v1}.

\bibitem[{O'Gorman and Schneider(2009)O'Gorman, and Schneider}]{OGorman2009}
O'Gorman, P.~A., and T.~Schneider, 2009: {The physical basis for increases in
  precipitation extremes in simulations of 21st-century climate change}.
  \textit{Proceedings of the National Academy of Sciences}, \textbf{106},
  14\,773--14\,777, \doi{10.1073/pnas.0907610106}.

\bibitem[{Pfahl et~al.(2017)Pfahl, O'Gorman,, and Fischer}]{Pfahl2017}
Pfahl, S., P.~A. O'Gorman, and E.~M. Fischer, 2017: {Understanding the regional
  pattern of projected future changes in extreme precipitation}. \textit{Nature
  Climate Change}, \textbf{7}, 423--427, \doi{10.1038/NCLIMATE3287}.

\bibitem[{Sch{\"{a}}r et~al.(2016)}]{Schar2016}
Sch{\"{a}}r, C., and Coauthors, 2016: {Percentile indices for assessing changes
  in heavy precipitation events}. \textit{Climatic Change}, \textbf{137},
  201--216, \doi{10.1007/s10584-016-1669-2}.

\bibitem[{Simmons and Burridge(1981)Simmons, and Burridge}]{Simmons1981}
Simmons, A.~J., and D.~M. Burridge, 1981: {An energy and angular-momentum
  conserving vertical finite-difference scheme and hybrid vertical
  coordinates}. \textit{Monthly Weather Review}, \textbf{109}, 758--766,
  \doi{10.1175/1520-0493(1981)109<0758:AEAAMC>2.0.CO;2}.

\bibitem[{Singh and O'Gorman(2012)Singh, and O'Gorman}]{Singh2012}
Singh, M.~S., and P.~A. O'Gorman, 2012: {Upward shift of the atmospheric
  general circulation under global warming: theory and simulations}.
  \textit{Journal of Climate}, \textbf{25}, 8259--8276,
  \doi{10.1175/JCLI-D-11-00699.1}.

\bibitem[{Stone(1968)}]{Stone1968}
Stone, H.~L., 1968: {Iterative solution of implicit approximations of
  multidimensional partial differential equations}. \textit{SIAM Journal on
  Numerical Analysis}, \textbf{5}, 530--558, \doi{10.1137/0705044}.

\bibitem[{Tandon et~al.(2018{\natexlab{a}})Tandon, Nie,, and
  Zhang}]{Tandon2018a}
Tandon, N.~F., J.~Nie, and X.~Zhang, 2018{\natexlab{a}}: {Strong influence of
  eddy length on boreal summertime extreme precipitation projections}.
  \textit{Geophysical Research Letters}, \textbf{45}, 10\,665--10\,672,
  \doi{10.1029/2018GL079327}.

\bibitem[{Tandon et~al.(2018{\natexlab{b}})Tandon, Xuebin,, and
  Sobel}]{Tandon2018}
Tandon, N.~F., Z.~Xuebin, and A.~H. Sobel, 2018{\natexlab{b}}: {Understanding
  the dynamics of future changes in extreme precipitation intensity}.
  \textit{Geophysical Research Letters}, \textbf{45}, 2870--2878,
  \doi{10.1002/2017GL076361}.

\bibitem[{Taylor et~al.(2012)Taylor, Stouffer,, and Meehl}]{Taylor2012}
Taylor, K.~E., R.~J. Stouffer, and G.~A. Meehl, 2012: {An overview of CMIP5 and
  the experiment design}. \textit{Bulletin of the American Meteorological
  Society}, \textbf{93}, 485--498, \doi{10.1175/BAMS-D-11-00094.1}.

\bibitem[{Zedan and Schneider(1983)Zedan, and Schneider}]{Zedan1983}
Zedan, M., and G.~E. Schneider, 1983: {A three-dimensional modified strongly
  implicit procedure for heat conduction}. \textit{AIAA Journal}, \textbf{21},
  295--303, \doi{10.2514/3.8068}.

\end{thebibliography}

%

%

\end{document}


\title{Supporting Information for ``Response of Vertical Velocities in Extratropical Precipitation Extremes to Climate Change''}


\author{Ziwei Li $^{*}$, Paul O'Gorman \\ \bigskip \bigskip
Department of Earth, Atmospheric and Planetary Sciences, Massachusetts Institute of Technology, Cambridge, MA, USA\\ 
$^{*}$ziweili@mit.edu \bigskip}

\vspace*{.01 in}
\maketitle
\vspace{.12 in}

\begin{figure}[ht]
\centering
    \centering
    \includegraphics[width=1.0\textwidth]{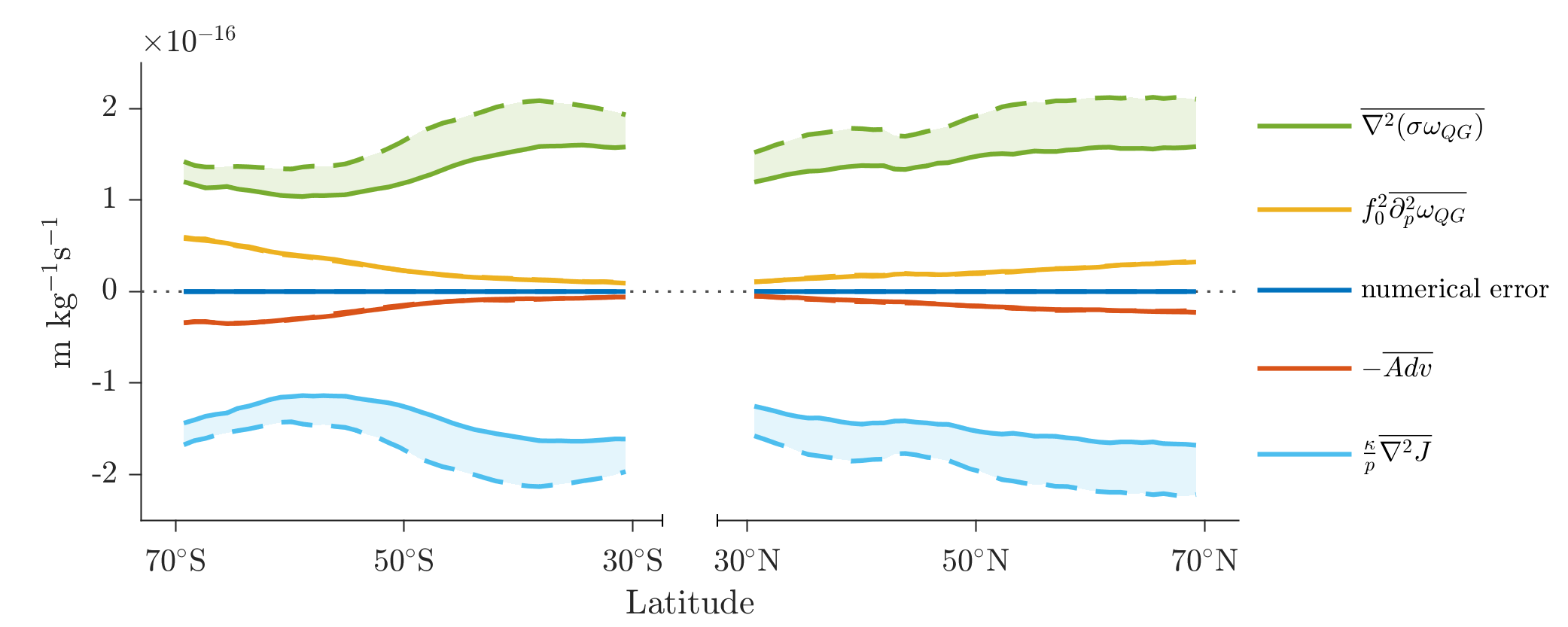}
\caption{
Event- and zonal-mean of terms in the QG-$\omega$ equation 
(Eq.\;1) 
at 500hPa for extratropical precipitation extremes in CESM-LE.
Solid lines indicate historical, dashed lines indicate RCP8.5, and shading indicates the response to climate change.
Terms on the right-hand side of the equation are shown with a minus sign so that the sum is zero.
The darker blue line gives the error in the numerical solution of the QG-$\omega$ equation.
}
\label{fig:terms_zonal}
\end{figure}

\begin{figure}[ht]
\centering
    \centering
    \includegraphics[width=0.7\textwidth]{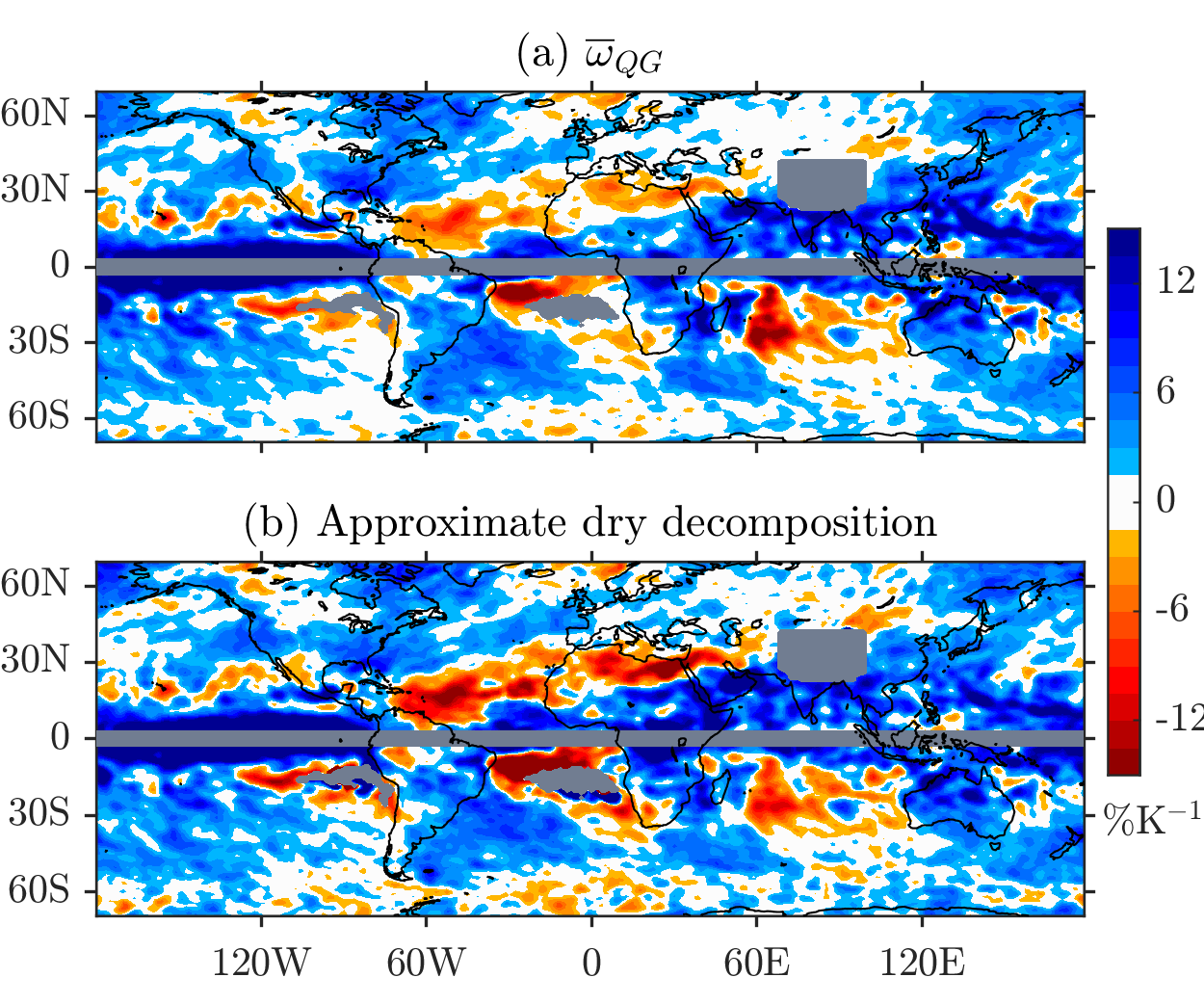}
\caption{
Changes in $\overline{\omega}_{\text{QG}}$ (a) from the numerical inversions
of the QG-$\omega$ equation and 
(b) as estimated from the approximate dry decomposition
(Eq.\;7) 
for vertical velocities at 500hPa associated
with precipitation extremes in CESM-LE. Results are expressed as
percentage changes relative to the historical climate and normalized
by the increase in global-mean surface temperature.
Masking and smoothing are as in 
Fig.\;2. 
}
\label{fig:approx_drydecomp}
\end{figure}

\begin{figure}[ht]
\centering
    \centering
    \includegraphics[width=1.0\textwidth]{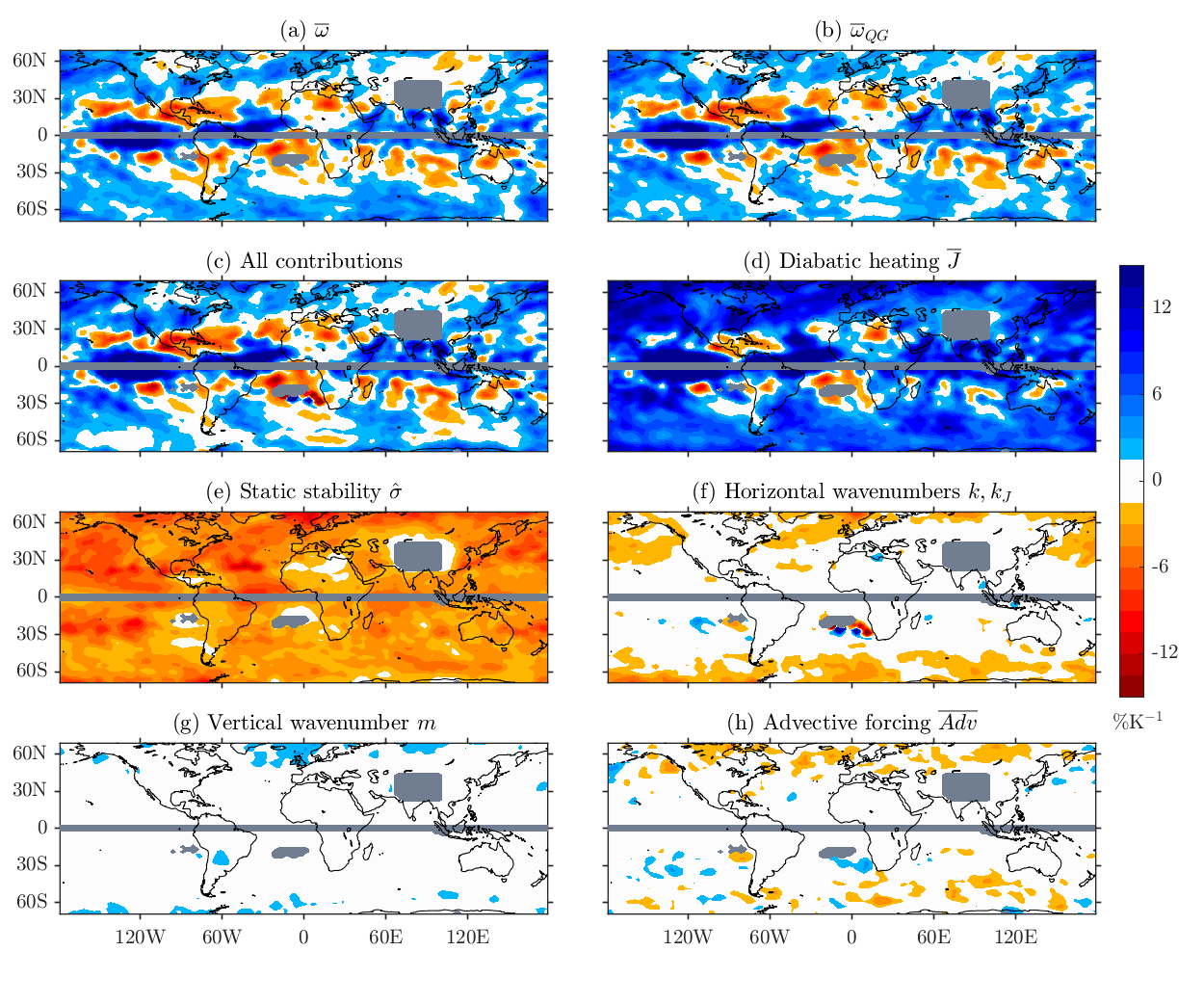}
\caption{As in 
Fig.\;4, 
but for 6-hourly events with GFDL-CM3. Grid points with fewer than 15 events are masked.}
\label{fig:drydecomp_GFDL}
\end{figure}

\begin{figure}
\centering
\includegraphics[width=1.0\textwidth]{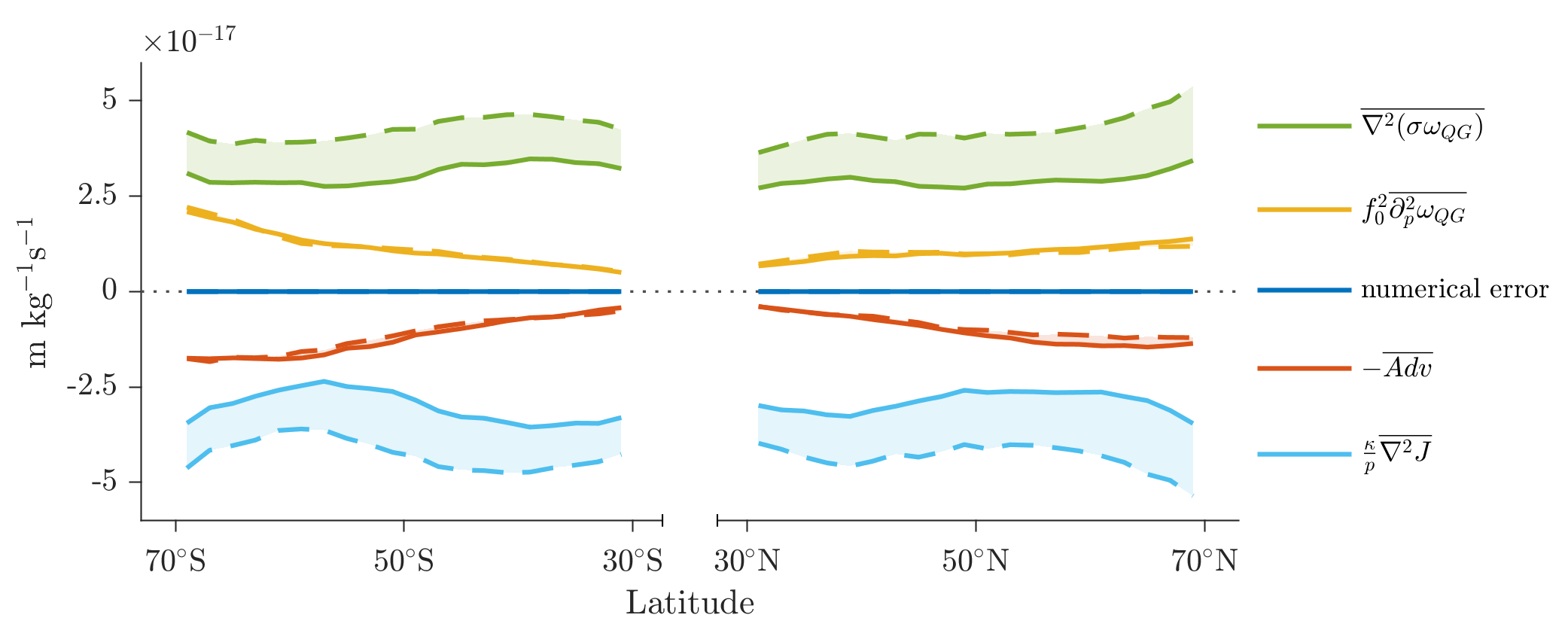}
\caption{As in Fig.\;\ref{fig:terms_zonal}, but for GFDL-CM3. (Note the change of scale of the vertical axis)}
\label{fig:terms_GFDL}
\end{figure}

\begin{figure}
\centering
\includegraphics[width=1.0\textwidth]{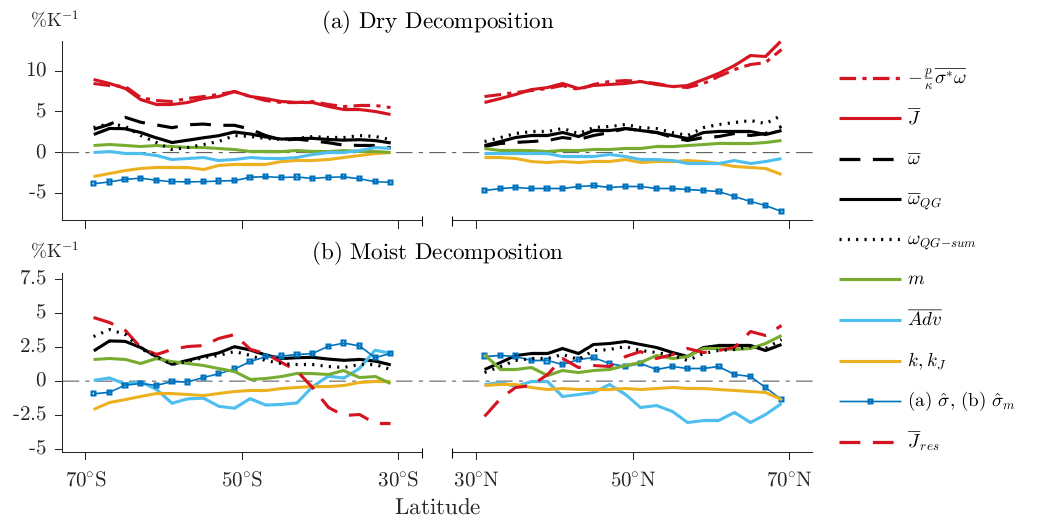}
\caption{As in 
Fig.\;5, 
but for GFDL-CM3. (Note the change of scale of the vertical axis)}
\label{fig:zonaldecomp_GFDL}
\end{figure}

\begin{figure}
\centering
\includegraphics[width=1.0\textwidth]{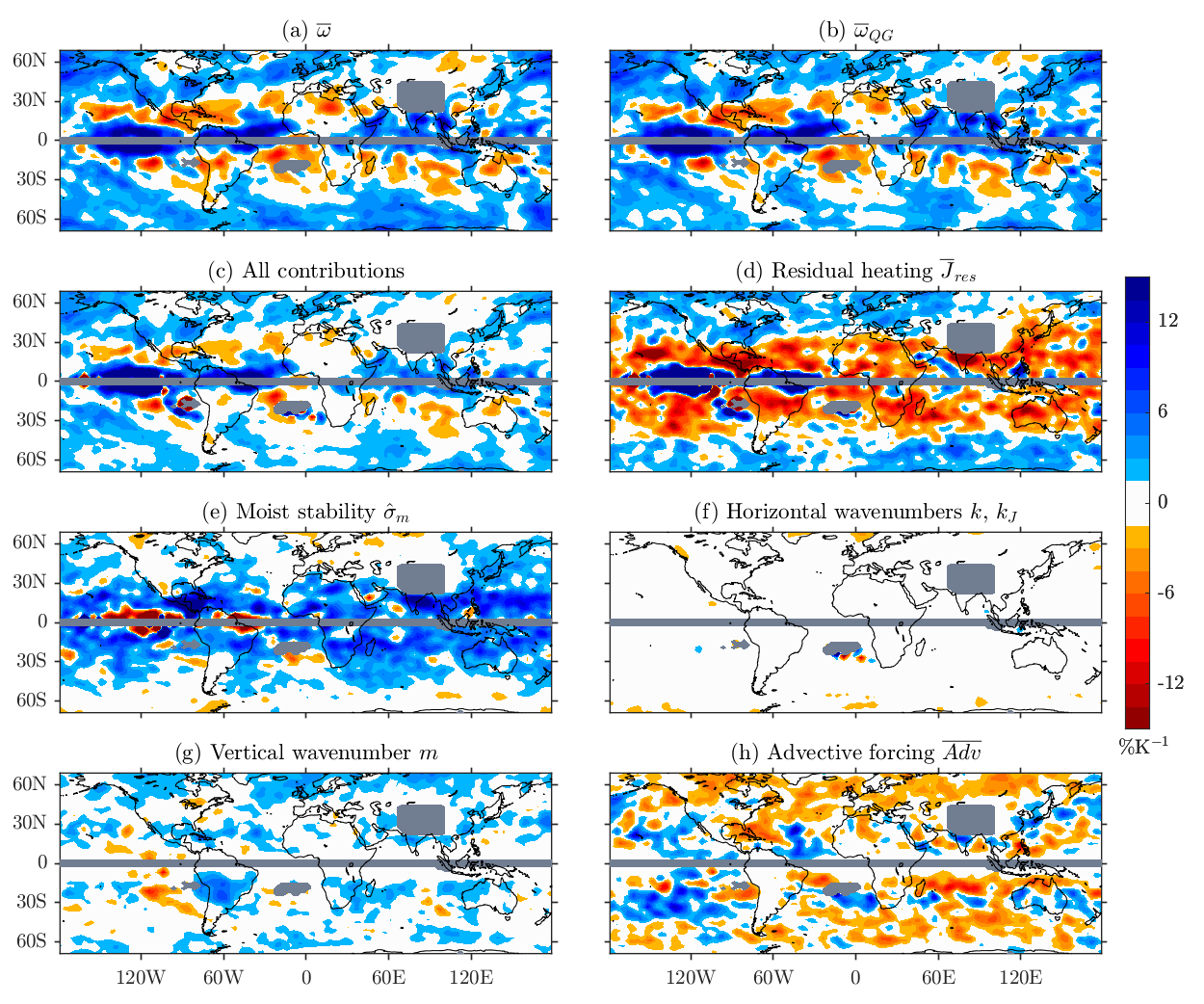}
\caption{As in 
Fig.\;8, 
but for GFDL-CM3.
}
\label{fig:moistdecomp_GFDL}
\end{figure}

\begin{figure}
\centering
\includegraphics[width=1.0\textwidth]{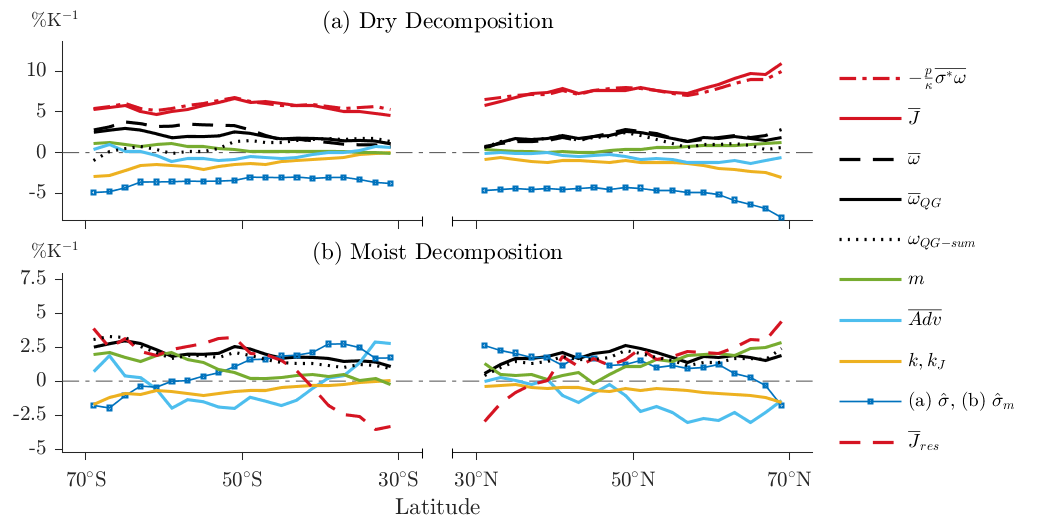}
\caption{As in 
Fig.\;5, 
but for GFDL-CM3 and the QG-$\omega$ equation is solved using $\omega$ taken from the GCM simulations for the lateral- and lower-boundary conditions.}
\label{fig:boundary_decomp}
\end{figure}

\begin{figure}
\centering
\includegraphics[width=1.0\textwidth]{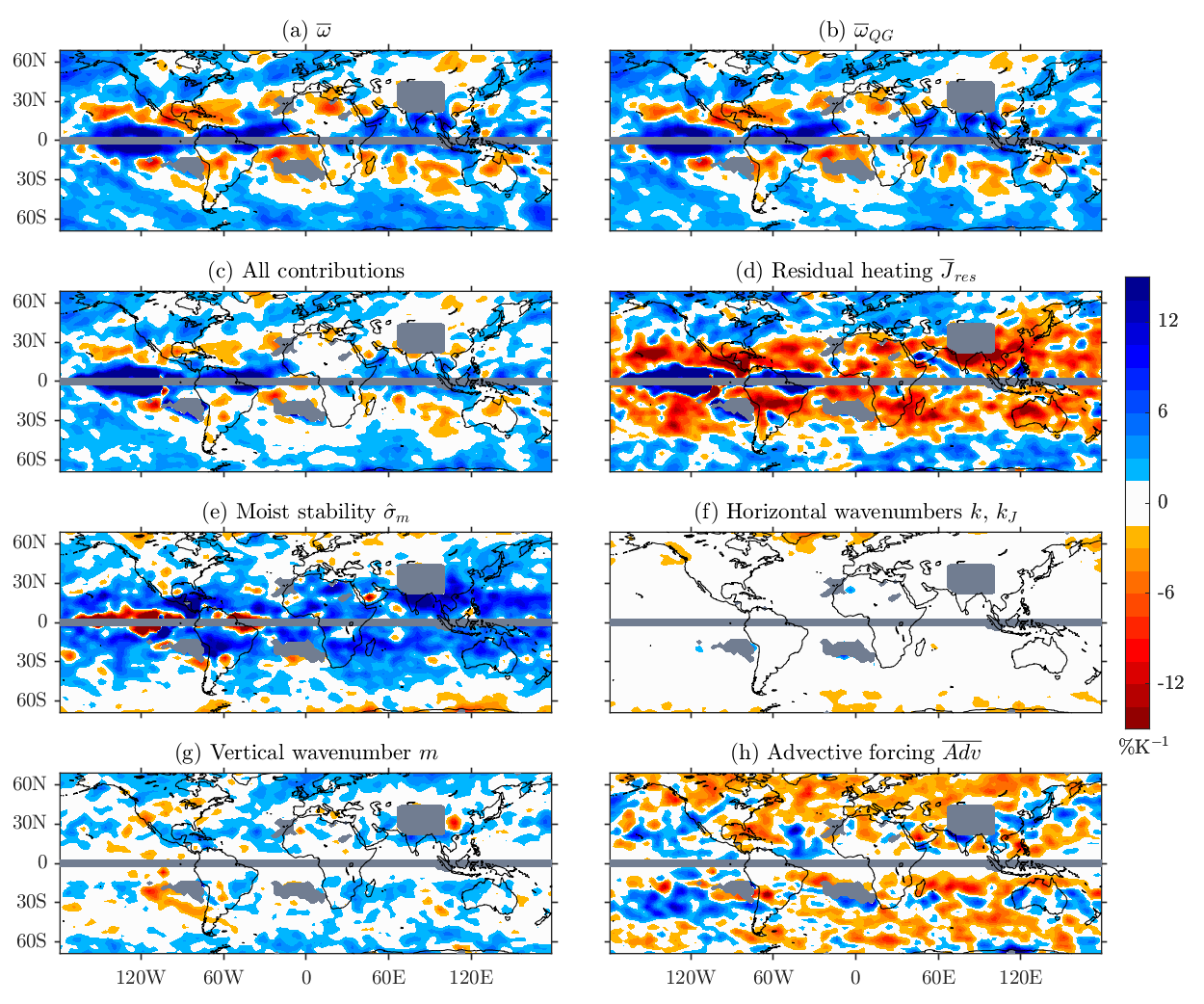}
\caption{As in 
Fig.\;8, 
but for GFDL-CM3 and the QG-$\omega$ equation is solved using $\omega$ taken from the GCM simulations for the lateral- and lower-boundary conditions.}
\label{fig:moistdecomp_GFDL_boundary}
\end{figure}


\begin{figure}
\centering
    \centering
    \includegraphics[width=1.0\textwidth]{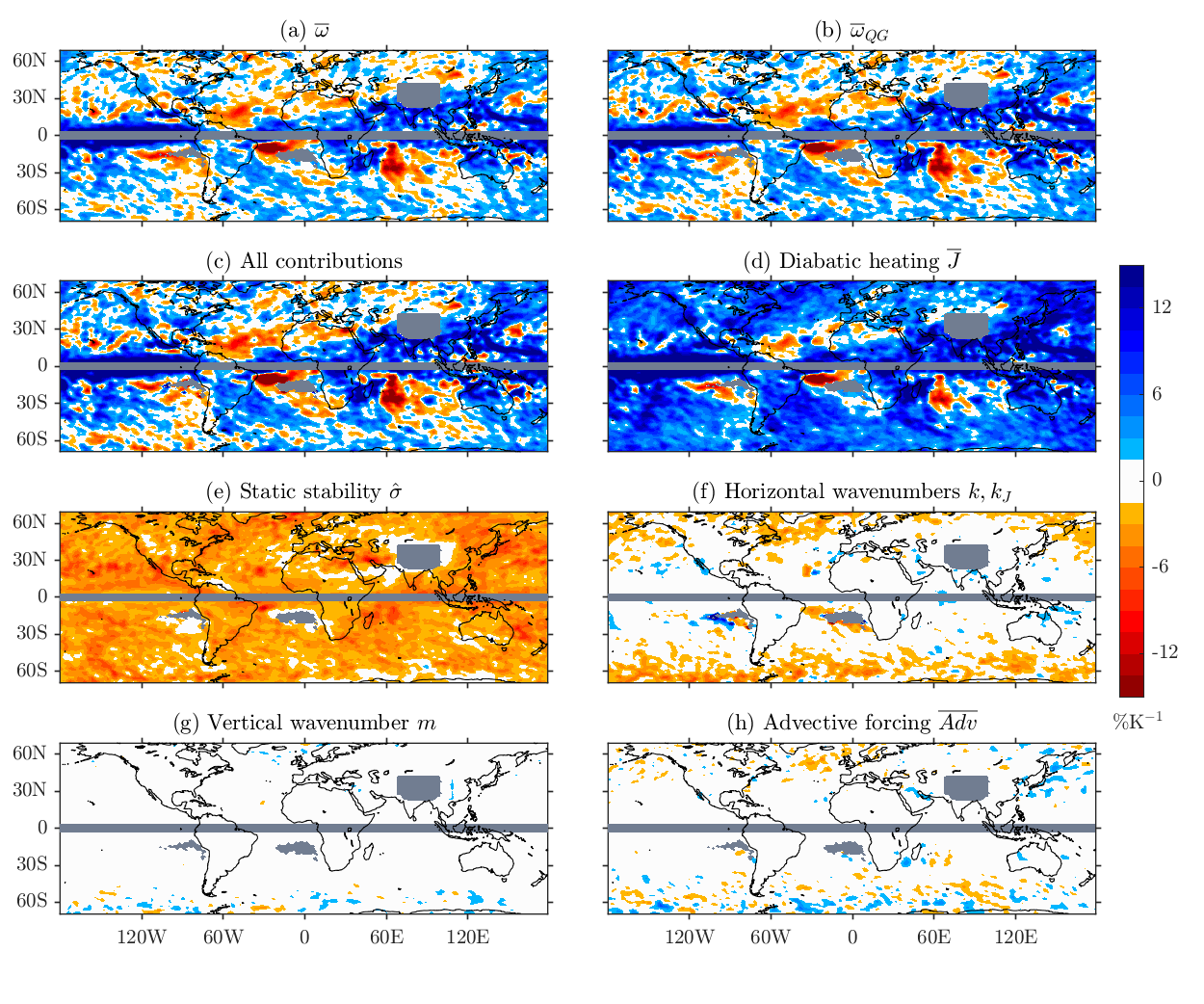}
\caption{As in 
Fig.\;4, 
but for daily precipitation extremes in CESM-LE, and grid points with fewer than 5 events are masked. }
\label{fig:drydecomp_CESM_daily}
\end{figure}

\begin{figure}
\centering
\includegraphics[width=1.0\textwidth]{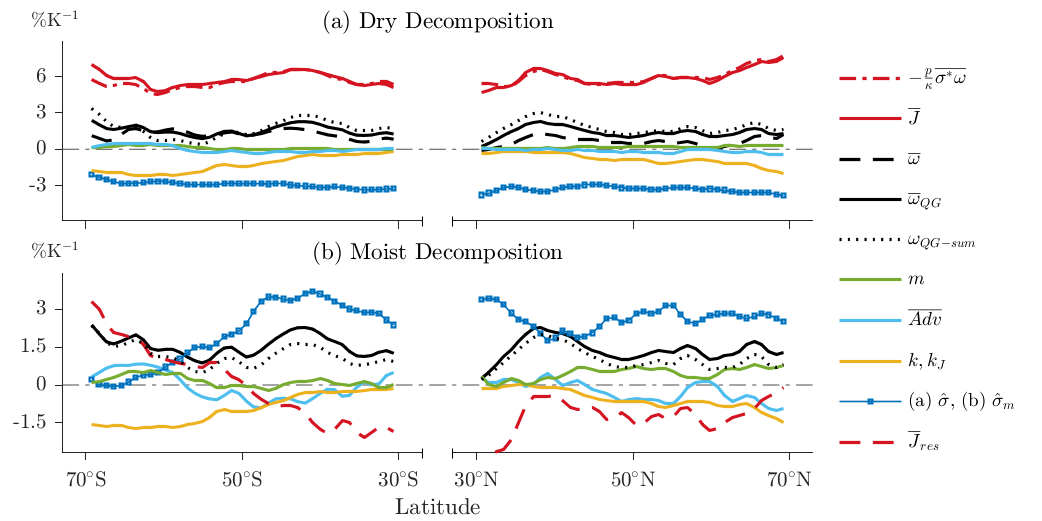}
\caption{As in 
Fig.\;5, 
but for daily precipitation extremes in CESM-LE.}
\label{fig:zonaldecomp_CESM_daily}
\end{figure}

\begin{figure}
\centering
    \centering
    \includegraphics[width=1.0\textwidth]{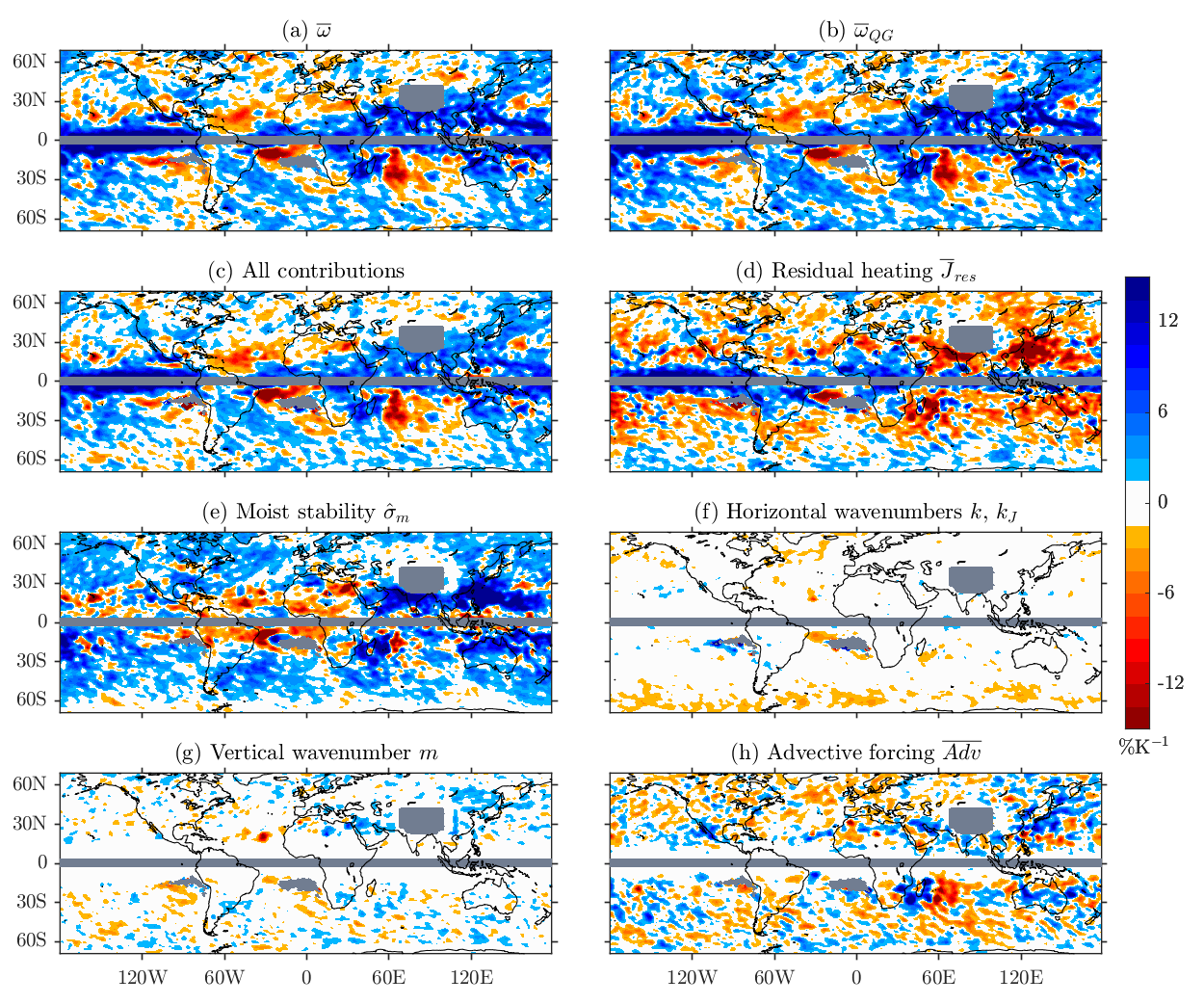}
\caption{As in 
Fig.\;8, 
but for daily precipitation extremes in CESM-LE, and grid points with fewer than 5 events are masked. }
\label{fig:moistdecomp_CESM_daily}
\end{figure}


\begin{figure}
\centering
    \centering
    \includegraphics[width=1.0\textwidth]{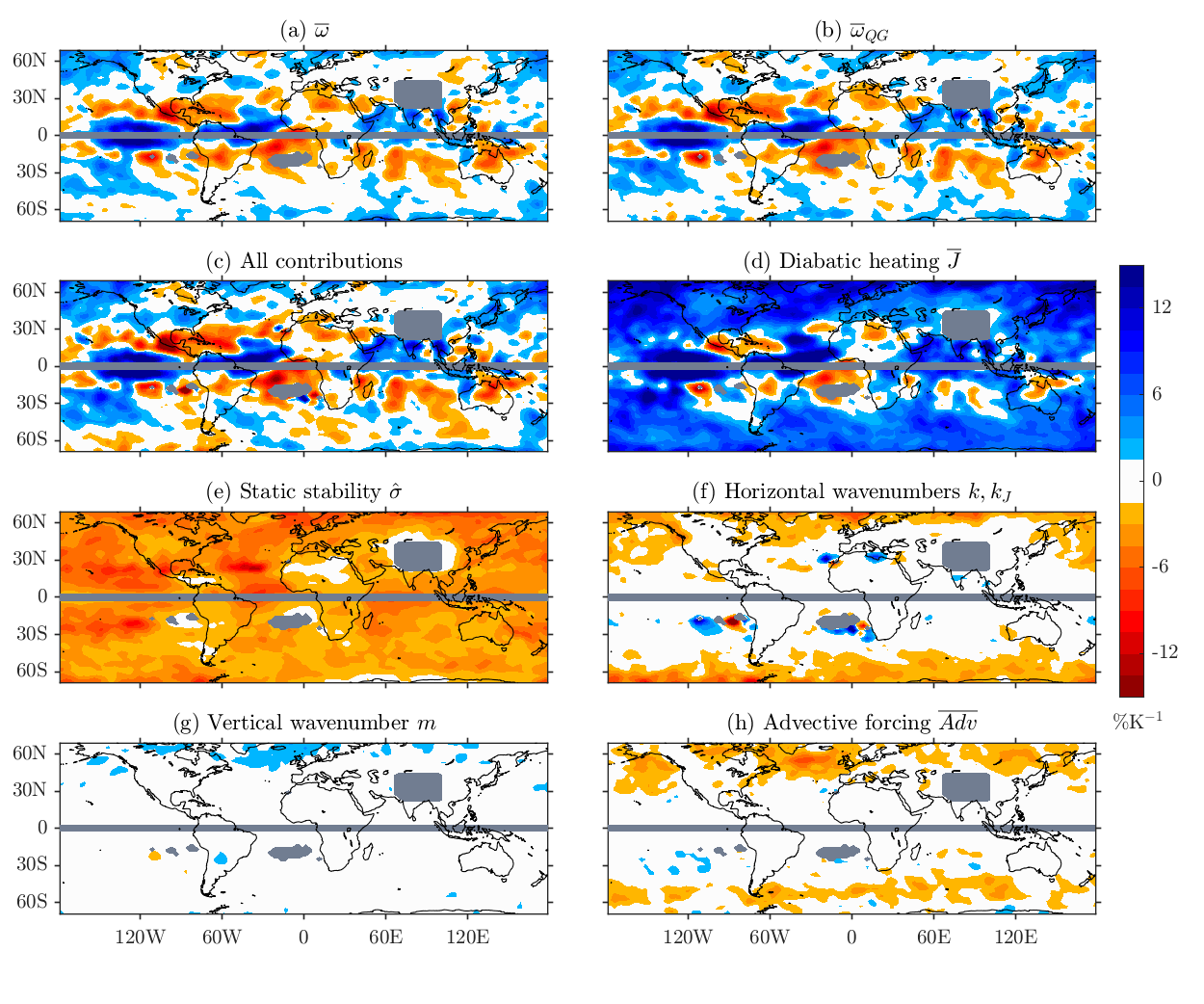}
\caption{As in 
Fig.\;4, 
but for daily precipitation extremes at the 99.5-percentile in GFDL-CM3, and grid points with fewer than 5 events are masked. }
\label{fig:drydecomp_GFDL_daily}
\end{figure}

\begin{figure}
\centering
\includegraphics[width=1.0\textwidth]{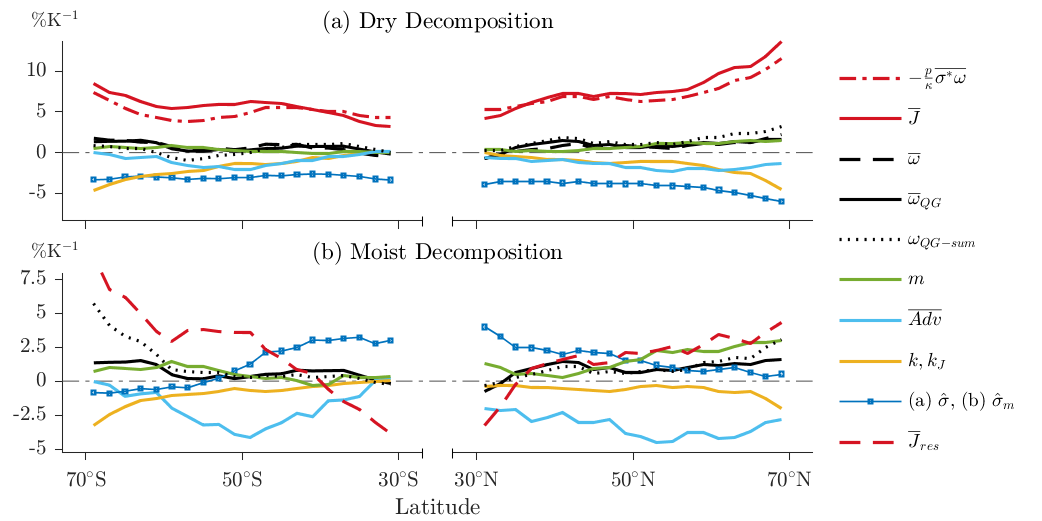}
\caption{As in 
Fig.\;5, 
but for daily precipitation extremes at the 99.5-percentile in GFDL-CM3.}
\label{fig:zonaldecomp_GFDL_daily}
\end{figure}

\begin{figure}
\centering
\includegraphics[width=1.0\textwidth]{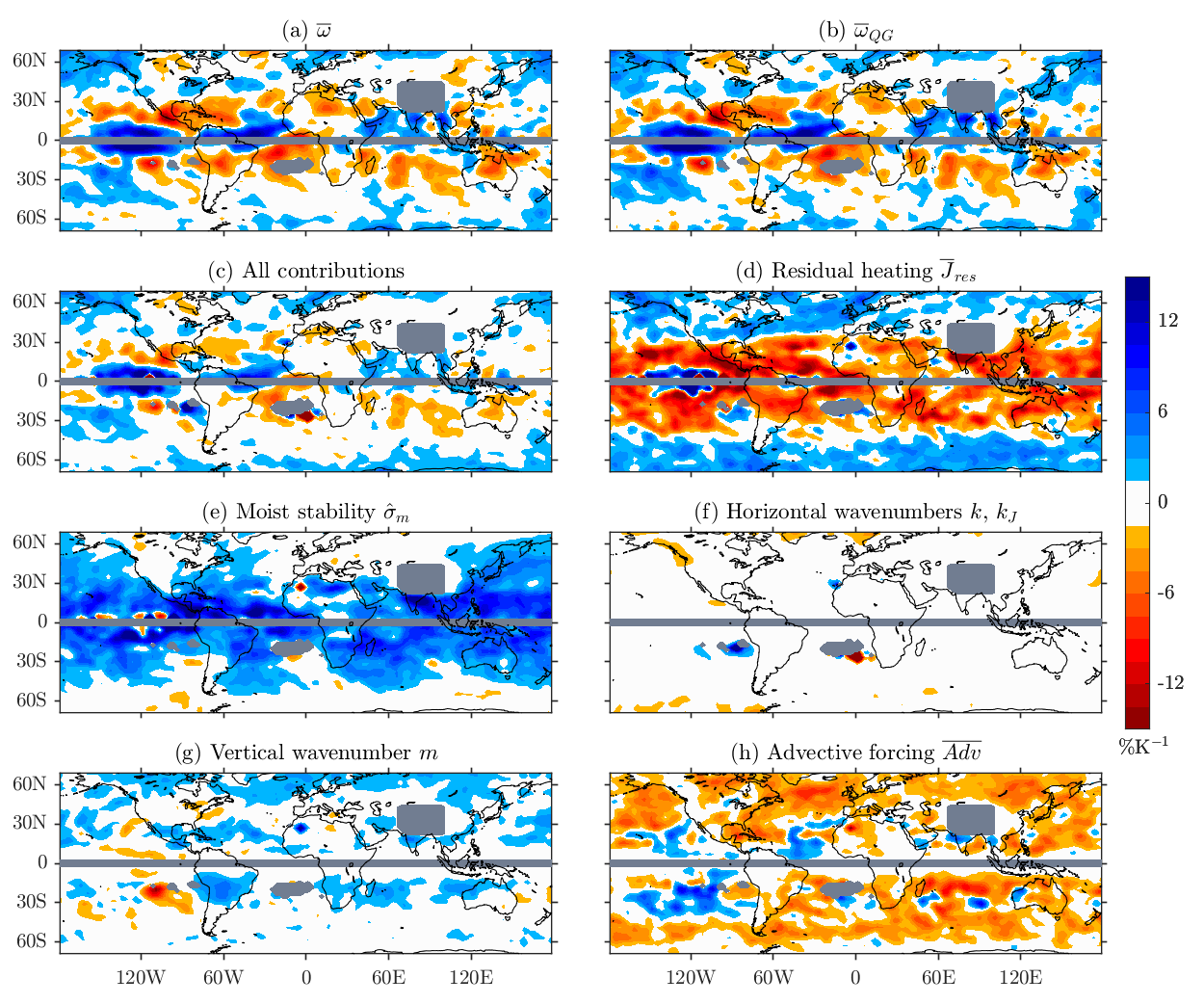}
\caption{As in 
Fig.\;8, 
but for daily precipitation extremes at the 99.5-percentile in GFDL-CM3.}
\label{fig:moistdecomp_GFDL_daily},
\end{figure}